\documentclass[fleqn,usenatbib]{mnras}

\usepackage[T1]{fontenc}
\usepackage{ae,aecompl}
\usepackage[normalem]{ulem}

\usepackage{macros_vdb} 

\usepackage{graphicx}	
\usepackage{amsmath}	
\usepackage{amssymb}	
\usepackage[dvipsnames]{xcolor}  
\usepackage{soul}
\usepackage{xcolor}
\usepackage[caption=false]{subfig}
\usepackage{color}

\title[The Mass Density of Free-Floating Black Holes]
{Constraining the mass density of free-floating black holes using razor-thin lensing arcs}

\author[Banik et al.]{%
   Uddipan Banik$^{1}$\thanks{E-mail: uddipan.banik@yale.edu},
   Frank~C.~van den Bosch$^1$,
   Michael Tremmel$^{2,3}$,
   Anupreeta More$^{4,5}$,
   \newauthor
   Giulia Despali$^{6}$,
   Surhud More$^{4,5}$,
   Simona Vegetti$^{6}$,
   and John P. McKean$^{7, 8}$
\vspace*{8pt}
\\
   $^1$Department of Astronomy, Yale University, PO. Box 208101, New Haven, CT 06520, USA\\
   $^2$Department of Physics, Yale University, PO Box 201820, New Haven, CT 06520, USA\\
   $^3$Yale Center for Astronomy and Astrophysics, PO Box 208121, New Haven, CT 06520, USA\\
   $^4$Kavli Institute for the Physics and Mathematics of the Universe (WPI), Todai Institutes of Advanced Study, University of Tokyo,\\5-1-5 Kashiwanoha, Kashiwa 277-8583, Japan\\
   $^5$The Inter-University Center for Astronomy and Astrophysics, Post Bag 4, Ganeshkhind, Pune, 411007, India\\
   $^6$Max Planck Institute for Astrophysics, Karl-Schwarzschild-Strasse 1, D-85740 Garching, Germany\\
   $^7$ASTRON, Netherlands Institute for Radio Astronomy, Postbus 2, NL-7990 AA, Dwingeloo, the Netherlands\\
   $^8$Kapteyn Astronomical Institute, PO Box 800, NL-9700 AV Groningen, the Netherlands\\
   }

\date{Accepted XXX. Received YYY; in original form ZZZ}

\pubyear{2018}

\begin{document}

\label{firstpage}
\pagerange{\pageref{firstpage}--\pageref{lastpage}}
\maketitle

\begin{abstract}
Strong lensing of active galactic nuclei in the radio can result in razor-thin arcs, with a thickness of less than a milli-arcsecond, if observed at the resolution achievable with very long baseline interferometry (VLBI). Such razor-thin arcs provide a unique window on the coarseness of the matter distribution between source and observer. In this paper, we investigate to what extent such razor-thin arcs can constrain the number density and mass function of `free-floating' black holes, defined as black holes that do not, or no longer, reside at the centre of a galaxy. These can be either primordial in origin or arise as by-products of the evolution of super-massive black holes in galactic nuclei. When sufficiently close to the line of sight, free-floating black holes cause kink-like distortions in the arcs, which are detectable by eye in the VLBI images as long as the black hole mass exceeds $\sim 1000$ Solar masses. Using a crude estimate for the detectability of such distortions, we analytically compute constraints on the matter density of free-floating black holes resulting from null-detections of distortions along a realistic, fiducial arc, and find them to be comparable to those from quasar milli-lensing. We also use predictions from a large hydrodynamical simulation for the demographics of free-floating black holes that are not primordial in origin, and show that their predicted mass density is roughly four orders of magnitude below the constraints achievable with a single razor-thin arc. 
\end{abstract}




\section{Introduction}
\label{sec:intro}

Strong gravitational lensing is a powerful tool to probe the distribution of matter in our Universe on a variety of scales and across a large range in redshift \citep[e.g.,][]{Schneider.etal.92, Treu.10}. Particularly powerful is the notion that strong gravitational lensing can be used to probe the coarseness of the matter distribution along the line of sight by looking for {\it distortions} of arcs, arclets, rings, or multiply imaged sources arising from the strong gravitational lensing of some source due to a much more massive object. These distortions come in the form of flux-ratio anomalies \citep[e.g.,][]{Mao.Schneider.98, Dalal.Kochanek.02, Mao.etal.04, Metcalf.05}, modified time-delays \citep[also known as the Shapiro delay,][]{Keeton.Moustakas.09, Mohammed.etal.15}, or distortions in extended arcs \citep[e.g.,][]{Koopmans.05, More.etal.09, Vegetti.Koopmans.09, Vegetti.etal.10, Vegetti.etal.12, Hezaveh.etal.16b, Birrer.etal.17}. 

A particularly exciting development has been the use of image distortions to probe the abundance of dark matter (sub)haloes on sub-galactic scales, which holds the potential to shed light on the nature of dark matter \citep[e.g.,][]{Li.etal.16, Li.etal.17, Hezaveh.etal.16a, Despali.etal.18, Bayer.etal.18, Vegetti.etal.18}. In general, detecting lower mass haloes requires higher sensitivity, which in turn implies higher spatial resolution, or more small-scale structure in the source. Hence, one ideally wants to use arcs and Einstein rings that have been observed at the highest possible resolution. Over the last few years, state-of-the-art observations at radio and sub-millimetre wavelengths, with very long baseline interferometry (VLBI) and with the long baselines of the Atacama Large (sub)Millimetre Array (ALMA), respectively, have revealed a number of extremely-thin lensing arcs, with a thickness of around 2-25 mas 
\citep[e.g.,][]{ALMA.partnership.15,Wong.etal.17,Spingola.etal.18}. These are ideally suited to probe the presence of low-mass perturbers ($10^5 \Msun \lta M \lta 10^8 \Msun$) along the line of sight between the source and the observer, and of low-mass substructure associated with the main lens. 

These perturbers may consist of a wide variety of objects, including dark matter (sub)haloes, dwarf galaxies, globular clusters, and black holes. Among these, dark matter (sub)haloes are of particular interest, as being able to constrain their number density over the mass range $10^5 \Msun \lta M \lta 10^8 \Msun$ holds the potential to constrain the nature of dark matter (in particular, the mass of a WIMP-like particle). However, dark matter (sub)haloes are also relatively diffuse objects, and unless they are relatively massive $(M \gta 10^{9} \Msun$)  their impact on the razor-thin arcs is only detectable through sophisticated image-analysis\footnote{In the case of interferometric data, this is best done in the UV-plane.} such as in the studies by \citet{Vegetti.etal.10,Vegetti.etal.12, Vegetti.etal.14}, \citet{Hezaveh.etal.16b} and \citet{Birrer.etal.17}. In this paper we therefore focus on another type of perturber, namely black holes. Due to their extreme compactness, they cause a maximal, and therefore most easily detectable, distortion for a given mass. Throughout we focus on free-floating black holes, which we define as black holes that are well-separated from the stellar bodies of galaxies, and we investigate the relation between the number density of such free-floating black holes and the probability of detecting one or more distortions along razor-thin gravitational lensing arcs. In particular, we restrict ourselves to distortions that are easily detectable `by eye' in the images inferred from the VLBI data without sophisticated analysis. As we demonstrate below, such distortions are kink-like in shape. 

There are different mechanisms that can give rise to free-floating black holes. On the one hand, they can be `primordial' in origin and be generated by one of three mechanisms:  through some form of cosmological phase transition \citep[e.g.,][]{Hawking.etal.82, Kodama.etal.82}, through a temporary softening of the equation of state \citep[e.g.,][]{Jedamzik.97}, or through the collapse of large inhomogeneities \citep[][]{Carr.Lidsey.93, Leach.etal.00}. Such primordial black holes (hereafter PBHs) are an intriguing candidate for the dark matter. However, very stringent constraints have been obtained on the mass density of PBHs from a wide variety of studies \citep[see][for a comprehensive review]{Carr.etal.16}, leaving little room for PBHs making up all of the dark matter, especially if the PBHs are massive ($M_{\rm BH} \gta 10^3 \Msun$). Nevertheless, even if such massive PBHs only provide a small fraction of the dark matter, they may have important consequences; in particular, they could act as seeds for the supermassive black holes (SMBHs) in galactic nuclei \citep[e.g.,][]{Carr.Silk.18}. At masses above $\sim 10^3\Msun$, the dominant constraint on the number density of PBHs comes from the cosmic microwave background (CMB). Massive PBHs will accrete matter prior to recombination, and the resulting radiation output would leave imprints on the spectrum and anisotropies of the CMB that have not been observed \citep[e.g.][]{Ricotti.etal.08}, ruling out that PBHs with masses in the range $10^3 \Msun \lta M_{\rm BH} \lta 10^{11} \Msun$ contribute more than a fraction $10^{-5}$ of the dark matter \citep[but see][for more conservative constraints]{AliHaimoud.Kamionkowski.17}. \citet{Inoue.Kusenko.17} have also obtained upper bounds on the PBH abundance from the observation of X-ray compact objects in galaxies in the mass range of $1 \Msun \lta M_{\rm BH} \lta 2\times 10^7 \Msun$.

However, there is an alternative formation mechanism for free-floating BHs in the mass range, $10^3 \Msun \lta M_{\rm BH} \lta 10^{11} \Msun$ as mentioned above, which operates well after recombination, thereby evading the CMB constraints. This mechanism is a byproduct of the galaxy formation and evolution process. It is well known that galaxies harbour SMBHs in their centres with a mass that is tightly correlated with the velocity dispersion of the stellar body \citep[][]{Tremaine.etal.2002}. Due to the hierarchical nature of structure formation, galaxies merge, during which the SMBHs of the progenitors sink to the centre of the merger remnant, where they form a SMBH-binary. If a new merger occurs before the binary has coalesced, this merger scenario may give rise to SMBH-triplets \citep[e.g.,][]{Deane.etal.14}. The three-body interaction of such a triplet can result in the ejection of one of the SMBHs, which can thus become unbound and free-floating. In addition, the coalescence of a binary SMBH can  result in a velocity kick, which can be sufficiently large as to unbind the resulting SMBH remnant from a low-mass galaxy \citep[][]{Favata.etal.2004, Gonzalez.etal.2007}. Finally, the tidal forces acting on satellite galaxies as they orbit their host halo may strip them apart, resulting in free-floating black holes orbiting the central galaxy. If the orbit is sufficiently far from the galactic centre, or the halo has a substantially dense core, dynamical friction from the host halo can be sufficiently small that such free-floating BHs survive for longer than the Hubble time \citep[e.g.,][]{DiCintio.etal.2017, Tremmel.etal.18A, Tremmel.etal.18}. In what follows we shall refer to these free-floating black holes that form as a by-product of galaxy formation as `wandering' black holes, in order to distinguish them from the PBHs discussed above. 

The existence of wandering SMBHs far from the galactic centre has been predicted using both cosmological simulations \citep{Bellovary.etal.2010, Volonteri.etal.16} and semi-analytic models \citep{Volonteri&Perna.2005}. Only recently have large-scale cosmological simulations been able to accurately follow the dynamics of SMBHs within galaxies down to sub-kpc scales \citep{Tremmel.etal.15, Tremmel.etal.17}. In particular, \citet{Tremmel.etal.18} use data from the {\sc Romulus25} cosmological simulation to predict that wandering SMBHs should be common-place in Milky Way-mass haloes at $z=0$, with $\sim 10$ existing within the virial radius. 
 
As is evident from the discussion above, constraining the number density and mass function of free-floating black holes can put powerful constraints on both inflationary models and the various physical mechanisms at play during the formation and evolution of SMBHs. Free-floating BHs are located in regions with little gas and/or stars, and they are therefore unlikely to reveal their presence through the emission associated with the accretion of matter. However, they can reveal their presence through the gravitational distortion of (razor-thin) lensing arcs, which is the phenomenon we investigate in this paper. Interestingly, as this paper was close to completion, \citet{Chen.etal.18} reported a possible detection of an SMBH of mass $8.4^{+4.3}_{-1.8} \times 10^9 \Msun$ offset by $4.4 \pm 0.3$ kpc from the centre of the main lensing galaxy, the brightest cluster galaxy of MACS J1149+2223.5 at $z=0.54$. The presence of a SMBH is inferred through a kink-like distortion in one of the multiply-lensed images of the background source. Although other explanations for the observed structure are possible, such a detection would be a wonderful proof of concept for the methodology advocated here.

This paper is organized as follows. In Section \ref{sec:double} we discuss the double lens configuration comprising a dark matter halo as the main lens plus a black hole along the line of sight, which acts as a secondary lens or a perturber distorting the arc produced by the main lens. In Section~\ref{sec:detectability} we discuss the criteria under which the presence of such a black hole is detectable as a perturbation of the lensing arc,  which we use in Section~\ref{sec:Veff} to compute constraints on the comoving number density of black holes, given some detection, or lack thereof. Section~\ref{sec:massdensityconstraint} discusses the kind of constraints that are realistically achievable, and we summarize our findings in Section~\ref{sec:concl}. We also provide an Appendix in which we compare the lensing distortion effect of BHs to that of subhaloes as well as derive a number of useful scaling relations showing the dependence of the constraints on the mass density of free-floating black holes on the black hole mass and the spatial resolution of the data.

Throughout this paper, we adopt the Planck 2014 cosmology \citep{Planck.14} with $H_0=67.8$~$\mathrm{km\,s^{-1}~ {Mpc}^{-1}}$, $\Omega_{\mathrm{DM,0}}=0.259$, $\Omega_{\mathrm{m0}}=0.307$ and $\Omega_{\Lambda,0}=0.693$. 

\section {The double lens configuration}
\label{sec:double}

We examine the distortion of a (razor-thin) lensing arc due to the presence of a perturbing black hole along the line of sight. We refer to the massive, primary lens that gives rise to the arc as the lens (L), and to the black hole as the perturber (P). If P is located sufficiently close to the geodesic connecting source (S) and observer (O), its gravitational lensing can cause a significant, localized perturbation of the arc, which is the signature we are considering here. In particular, we aim to compute the effective volume, centred around this geodesic, inside of which a perturbing black hole of a given mass, $M_\rmP$, causes such a detectable distortion.
\begin{figure*}
\centering
\hspace{-2mm}
\subfloat[][Foreground\label{fig1a}]{\includegraphics[width=0.75\textwidth,height=0.45\textwidth]{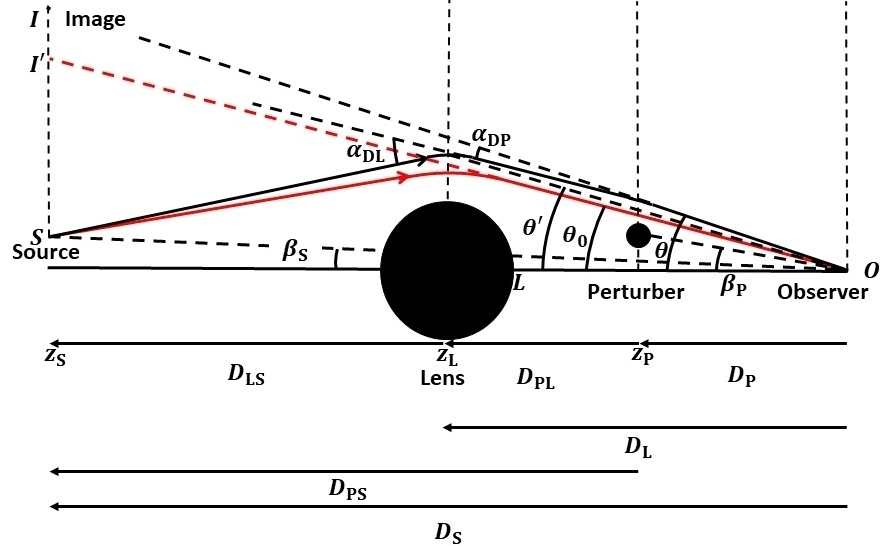}} \\
\vspace{-0.5mm}
\subfloat[][Background\label{fig1b}]{\includegraphics[width=0.75\textwidth,height=0.45\textwidth]{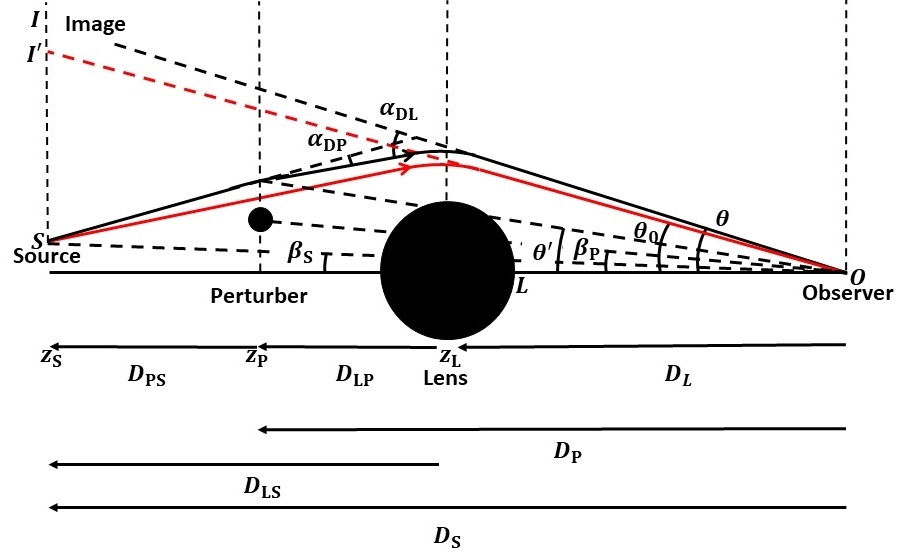}}
\caption{\label{fig1} Schematic of the foreground and background double lensing configurations as viewed from the side. The black and red rays denote the perturbed and unperturbed geodesics respectively. Only one component of the angles is shown.}                                     
\end{figure*}                                               

The lens and the perturber are generally located at different redshifts. If they reside at the same redshift, the overall deflection of light from the source due to the combined effect of the two lenses is simply the sum of the deflections caused by L and P. However, in the case that the two are located at different redshifts, the overall angle of deflection is determined by the double lens equation \citep[][]{Schneider.etal.92, Keeton.03}, which is the equation we use throughout. Note that the perturber can be either in the foreground (between the lens and observer), or in the background (between the source and lens).  Both configurations are depicted in Fig.~\ref{fig1}, which specifies the various angles and distances used throughout.

We use $D_\rmS$, $D_\rmL$ and $D_\rmP$ to refer to the (physical) angular diameter distances from the observer to the source, the lens and the perturber, respectively. $D_{\rm PS}$ and $D_{\rm LS}$ indicate the angular diameter distances from P to S, and L to S, respectively. Here
\begin{equation}
D_{12} \equiv \frac{r_{12}}{1+z_2}
\end{equation}
where $r_{12}$ is the comoving distance between two objects along the same line of sight located at redshifts $z_1$ and $z_2 > z_1$ \citep[e.g.,][]{MBW10}. Finally, $D_{\rm PL}$ indicates the angular diameter distance from the perturber to the lens in the foreground configuration (depicted in Fig.~\ref{fig1a}) while $D_{\rm LP}$ stands for the same from the lens to the perturber in the background (depicted in Fig.~\ref{fig1b}) configuration.

\subsection{Foreground configuration}
\label{sec:foreground}

First let us consider the foreground configuration depicted in Fig.~\ref{fig1a}.  Let $\vec{\theta}$ be the angle between the image, I, and the line OL (hereafter `baseline'), joining the main lens (located at redshift $z_\rmL$) and observer. Let $\vec{{\alpha}_{\rm{DL}}}$ and $\vec{{\alpha}_{\rm{DP}}}$ be the angles of deflection due to the lens and the perturber subtended at the lens and the perturber, respectively, and let $\vec{{\alpha}_{\rm{L}}}$ and $\vec{{\alpha}_{\rm{P}}}$ be the corresponding deflection angles subtended at the observer. These are related according to
\begin{equation}
\vec{\alpha_\rmL} = \frac{D_{\rm LS}}{D_\rmS} \, \vec{\alpha_{\rm DL}}\,, 
\;\;\;\;\;\;\;\;\;\;\;\;\;\;\;\;
\vec{\alpha_\rmP} = \frac{D_{\rm PS}}{D_\rmS} \, \vec{\alpha_{\rm DP}}\,.
\end{equation}
Finally, we define $\vec{{\beta}_\rmS}$ and $\vec{{\beta}_\rmP}$ as the angles between the baseline, OL, and the lines-of-sight towards S and P, respectively.

When the perturber is present, the double lens equation describing the system can be written as
\begin{equation}
\vec{\beta_\rmS}=\vec{\theta} - \vec{\alpha_\rmP}(\vec{\theta}) - \vec{\alpha_\rmL}(\vec{\theta'})\,.
\label{perturb}
\end{equation}
Here
\begin{equation}
\vec{\theta'} = \vec{\theta} - \gamma_\rmf \, \vec{\alpha_\rmP} (\vec {\theta})
\end{equation}
with
\begin{equation}
\gamma_\rmf \equiv \frac{D_{\rm{PL}}}{D_\rmL}\frac{D_\rmS}{D_{\rm{PS}}} \,.
\end{equation}

In absence of the perturber, the lens equation reduces to
\begin{equation}
\begin{aligned}
\vec{\beta_\rmS}=\vec{\theta_0} - \vec{\alpha_\rmL}(\vec{\theta_0})\,,
\end{aligned}
\label{unperturb}
\end{equation}
where $\vec{\theta_0}$ is the image angle in absence of the perturber. 

Subtracting equation~(\ref{perturb}) from equation~(\ref{unperturb}) and expanding perturbatively to linear order around $\vec{\theta_0}$, we obtain
\begin{equation}
\begin{aligned}
\delta \vec{\theta} \equiv \vec{\theta} - \vec{\theta_0} = {\left(1-\vec{\nabla_{\theta}}\vec{\alpha_\rmL}(\vec{\theta_0})\right)}^{-1} \left(1-\gamma_\rmf \vec{\nabla_{\theta}}\vec{\alpha_\rmL}(\vec{\theta_0}) \right) \vec{\alpha_\rmP} (\vec{\theta})\,.
\label{deltheta}
\end{aligned}
\end{equation}
This perturbative form of the lens equation can be solved numerically\footnote {To numerically solve the perturbative form of the lens equation we use the SciPy \citep{Jones.etal.09} module {\tt fsolve}, which is a wrapper around MINPACK's {\tt hybrd} and {\tt hybrj} algorithms. Both find the roots of a system of $N$ non-linear equations with $N$ variables using a modified form of the Powell hybrid method \citep{Powell.1970}.} for the distortion, $\delta \vec{\theta}$, due to the perturber $P$.

\subsection{Background configuration}
\label{sec:background}

If the perturbing black hole is located in the background (i.e., between the primary lens and the source; see Fig.~\ref{fig1b}), the double lens equation is of the form
\begin{equation}
\vec{\beta_\rmS}=\vec{\theta} - \vec{\alpha_\rmL}(\vec{\theta}) - \vec{\alpha_\rmP}(\vec{\theta'})\,.
\label{backperturb}
\end{equation}
Here
\begin{equation}
\vec{\theta'}=\vec{\theta} - \gamma_\rmb \vec{\alpha_\rmL} (\vec {\theta})\,,
\end{equation}
with 
\begin{equation}
\gamma_\rmb \equiv \frac{D_{\rm{LP}}}{D_\rmP}\frac{D_\rmS}{D_{\rm{LS}}}\,.
\end{equation}
Subtracting equation~(\ref{backperturb}) from equation~(\ref{unperturb}) and expanding perturbatively to linear order around $\vec{\theta_0}$, now yields
\begin{equation}
\begin{aligned}
\delta \vec{\theta} = {\left(1-\vec{\nabla_{\theta}}\vec{\alpha_\rmL}(\vec{\theta_0})\right)}^{-1} \vec{\alpha_\rmP} (\vec{\theta'})\,.
\label{backdeltheta}
\end{aligned}
\end{equation}

\subsection{Matrix notation}

It is useful to write the above expressions for the distortion $\delta \vec{\theta}$ in matrix notation:
\begin{align}\label{matrixnot}
\delta \vec{\theta} = \begin{cases}
\textbf{M}(\vec{\theta_0}) \, \textbf{C}(\vec{\theta_0}) \, \vec{\alpha_\rmP}(\vec{\theta})\,, & \text{foreground,} \\
\textbf{M}(\vec{\theta_0}) \, \vec{\alpha_\rmP}(\vec{\theta'})\,, & \text{background.} \\
\end{cases}
\end{align}
Here $\textbf{M} (\vec{\theta_0}) = {\left(1-\vec{\nabla_{\theta}}\vec{\alpha_\rmL}(\vec{\theta_0})\right)}^{-1}$ is the magnification tensor for the lens, and $\textbf{C} (\vec{\theta_0}) = \left(1-\gamma_\rmf\vec{\nabla_{\theta}}\vec{\alpha_\rmL}(\vec{\theta_0}) \right)$ is the correction tensor for the double lens configuration.

Throughout we define a Cartesian basis in which the $y$-axis connects the perturber to the baseline, and the $x$-axis is perpendicular to both the baseline and the $y$-axis. In this basis, the $x$-component of $\vec{\beta_\rmP}$ is zero by construction, 
\begin{equation}
\delta \vec{\theta} = \begin{bmatrix}
\delta \theta_\rmx\\
\delta \theta_\rmy 
\end{bmatrix}\,,
\end{equation}
and the  magnification and correction tensors are given by
\begin{equation}
\begin{aligned}
\textbf{M} (\vec{\theta_0}) = \frac{1}{1-\frac{\theta_\rmE}{\theta_0}} \begin{bmatrix} 
1-\theta_\rmE \frac{\theta_{0\rmx}^2}{\theta_0^3} & -\theta_\rmE \frac{\theta_{0\rmx}\theta_{0\rmy}}{\theta_0^3} \\ 
-\theta_\rmE \frac{\theta_{0\rmx}\theta_{0\rmy}}{\theta_0^3} & 1-\theta_\rmE \frac{\theta_{0\rmy}^2}{\theta_0^3} 
\end{bmatrix}\,,
\end{aligned}
\label{M}
\end{equation}
and
\begin{equation}
\begin{aligned}
\textbf{C} (\vec{\theta_0}) = \begin{bmatrix} 
1-\gamma_\rmf\,\theta_\rmE \frac{\theta_{0\rmy}^2}{\theta_0^3} & \gamma_\rmf\,\theta_\rmE \frac{\theta_{0\rmx}\theta_{0\rmy}}{\theta_0^3} \\ 
\gamma_\rmf\,\theta_\rmE \frac{\theta_{0\rmx}\theta_{0\rmy}}{\theta_0^3} & 1-\gamma_\rmf\,\theta_\rmE \frac{\theta_{0\rmx}^2}{\theta_0^3} 
\end{bmatrix}\,,
\end{aligned}
\label{C}
\end{equation}
where $\theta_0 \equiv |\vec{\theta_0}|$, and 
\begin{equation}
\theta_{\rmE} = \sqrt[]{\frac{2\pi G M_\rmL(\theta_\rmE)}{c^2}\frac{D_{\rm{LS}}}{D_{\rmL}D_{\rmS}}}
\label{thetaEinst}
\end{equation}
is the Einstein radius of the lens, with $M_\rmL(\theta_\rmE)$ being the 3D mass of the lens enclosed within a sphere of radius equal to the Einstein radius.

\begin{figure}
\includegraphics[width=0.48\textwidth,height=0.3\textwidth]{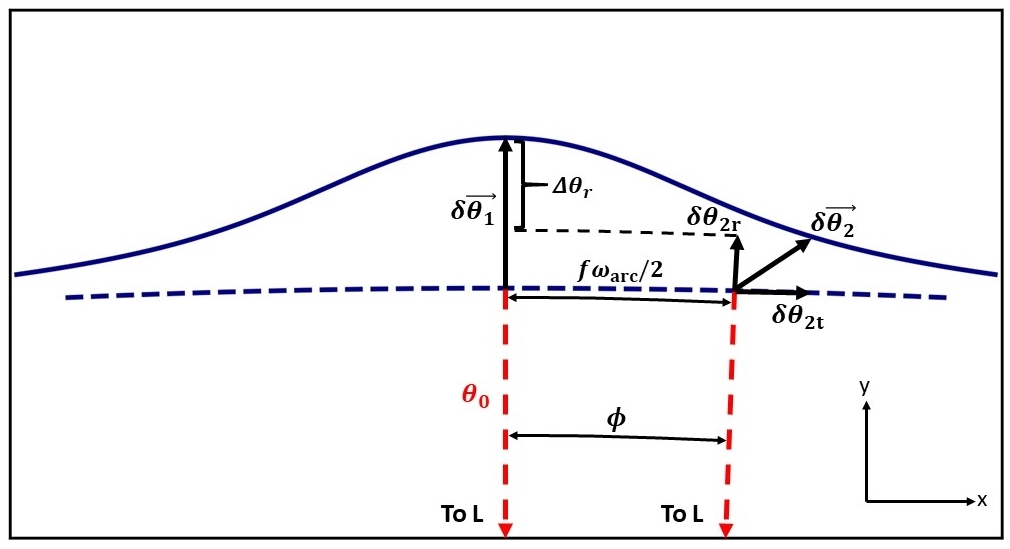}
\caption{Illustration of the various angles and angular perturbations involved in double lensing. The dashed and solid lines indicate the unperturbed and perturbed arcs, respectively. We consider the perturbation `detectable by-eye', if the difference $\Delta\theta_\rmr \equiv \delta\theta_{1{\rmr}} -
\delta\theta_{2{\rmr}}$ is sufficiently large compared to the spatial resolution of the observation
(see \S\ref{sec:detectability} for details). Here the subscript `1' refers to the point along the arc closest to the black hole, while subscript `2' refers to a point that is offset by an angle $f\omega_{\rm arc}/2$ along the arc, subtended from the observer. From the center of the lens, points 1 and 2 subtend an angle $\phi = \left(f \, \omega_{\rm arc}\right) / (2 \, |\vec{\theta_0}|)$.}
\label{angles_illustration}
\end{figure}

Although the Cartesian basis simplifies the above expressions for $\textbf{M}$ and $\textbf{C}$, in what follows we are mainly interested in the distortion, $\delta \theta_\rmr$, in the radial direction, where the radius $r$ is defined with respect to the position of the main lens. The perturbations in the radial and tangential directions follow from those in the $x$ and $y$-directions using a simple
rotation:
\begin{equation}
 \begin{bmatrix}
   \delta \theta_\rmr\\ 
   \delta \theta_\rmt
 \end{bmatrix}
 = 
 \begin{bmatrix}
   \sin\,\phi & \cos\, \phi\\
   \cos\, \phi & -\sin\,\phi
 \end{bmatrix} 
 \begin{bmatrix}
   \delta \theta_\rmx\\ 
   \delta \theta_\rmy
 \end{bmatrix}\,.
 \label{radtandeltheta}
\end{equation}
Here $\phi$ is the angle along the unperturbed arc subtended at the lens, where we define $\phi=0$ as the point closest to the perturber, i.e., where $(\delta\theta_\rmr, \delta\theta_\rmt) = (\delta\theta_y, \delta\theta_x)$. This is related to the angle 
$\omega$ along the arc subtended at the observer, according to  $\phi= \omega / |\vec{\theta}_0|$ (see Fig. \ref{angles_illustration}).

\subsection{Lens models}
\label{sec:lensmodels}

We model the primary lens as a singular isothermal sphere for which the angle of deflection is given by 
\begin{equation}
\vec{\alpha_\rmL}(\vec{\theta}) = \theta_\rmE \, \frac{\vec{\theta}}{|\vec{\theta}|}\,,
\label{lens}
\end{equation}
where $\theta_\rmE$ is the Einstein radius of the lens as defined in equation (\ref{thetaEinst}).

For simplicity, we assume that the source is extended but small enough that the thickness of the observed arc (or Einstein ring) mainly arises from broadening of the data due to the finite resolution of the observation. We also assume that the thickness is uniform along the arc. This effectively implies that we assume uniform sensitivity for detecting perturbers along the entire arc. In general, a single source-lens system can produce multiple arcs whenever the lens is not perfectly aligned with the source. If this is the case, we consider each arc as independent, having its own length and width. At the end of Section~\ref{sec:Veff} we briefly mention how to combine the constraints from multiple arcs.

Finally, the perturber, P, is modeled as a Schwarzschild black hole, which can be treated as a point mass. The deflection of a light ray passing at an impact parameter $b$ past a point mass $M_\rmP$ is given by $2 \, R_\rms/ b$. Here, the Schwarzschild radius $R_\rms=2 G M_\rmP / c^2$, where $G$ is the universal gravitational constant and $c$ is the speed of light. Hence, using that $b = D_\rmP \, |\vec{\theta} - \vec{\beta_\rmP}|$, the black hole's deflection angle subtended at the observer can be written as
\begin{equation}
\vec{\alpha_\rmP} = \theta_{\rm{EP}}^2 \, \frac{\vec{\theta} - \vec{\beta_\rmP}}{{\left|\vec{\theta} - \vec{\beta_\rmP}\right|}^2}\,,
\label{perturber}
\end{equation}
where $\theta_{\rm{\rm{EP}}}$ is the Einstein radius of the perturber, given by
\begin{equation}
\theta_{\rm {EP}} = \sqrt{\frac{2 \, R_\rms \, D_{\rm{PS}}}{D_{\rmP} \, D_{\rmS}}}\,.
\end{equation}
\begin{figure}
\includegraphics[width=0.52\textwidth,height=0.3\textwidth]{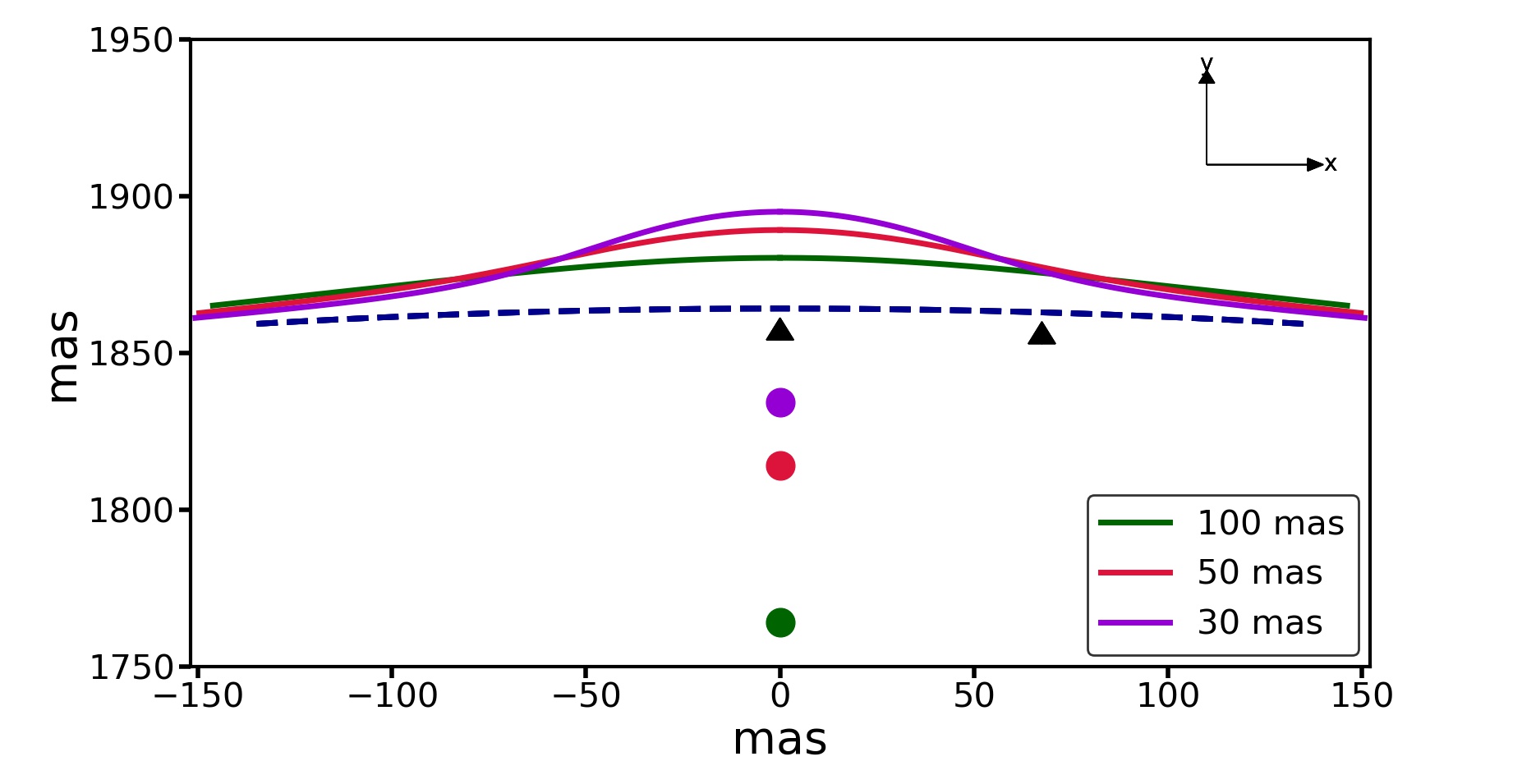}
\caption{Perturbed arc in presence of a $10^7 \Msun$ perturber at $z_\rmP=0.01$ for the fiducial case ($z_\rmL=0.881$, $z_\rmS=2.056$, $M_\rmL(\theta_\rmE) = 10^{12} \Msun$ and $|\vec{\beta_\rmS}|=36$ mas, i.e. $|\vec{\theta_0}| = 1.866"$), obtained by numerically solving the perturbative form of the double lens equation (\ref{matrixnot}). The blue dashed line indicates the unperturbed arc and the colored solid lines indicate the perturbed arcs for three different angular impact parameters of the perturber (green: $100$ mas, red: $50$ mas, violet: $30$ mas) defined with respect to the unperturbed arc. The black arrows indicate the points on the arc pertaining to the detectability criteria (see equation [\ref{detectcriteria}]). Note that for each angular impact parameter only one of the two perturbed arcs is shown here.}
\label{lensing_distortion}
\end{figure}

\subsection{Fiducial lensing configuration}
\label{sec:fiducial}

Throughout this paper, we consider a fiducial lensing configuration in which the razor-thin arc arises from a source at redshift $z_\rmS = 2.056$ that is being lensed by an isothermal sphere halo of mass $M_\rmL(\theta_\rmE) =10^{12} \Msun$ at a redshift $z_\rmL = 0.881$. We assume that the arc has a total length of $\arc = 270\,$mas, and has been observed with a spatial resolution of $\calR = 0.8\,$mas. Here, and throughout, $\calR$ is defined as the full-width half-maximum (FWHM) of the point-spread function, in the case of an optical image, or of the (synthesized) beam in the case of interferometric data. These numbers are motivated by an existing observation of a razor-thin arc imaged in the radio with VLBI for the gravitational lens JVAS~B1938+666 (McKean et al., in preparation). This spatial resolution is typical for VLBI imaging of gravitational lenses at 1.7 GHz with a global array \citep[e.g.][]{Spingola.etal.18}, although the numbers can change by a factor of a few depending on the actual {\it uv}-coverage of the observations, which is a function of the hour-angle and declination of the source.

In the strong lensing regime, where $|\vec{\beta_\rmS}| < \theta_\rmE$, $\vec{\theta_0}$ has two solutions
\begin{equation}
\vec{\theta_0} = \vec{\beta_\rmS} \pm \vec{\theta_\rmE}\,.
\end{equation}
Throughout we adopt $|\vec{\beta_\rmS}|=36$ mas, which is significantly less than the Einstein radius, $\theta_\rmE$, which is $1.83$ arcsec for our fiducial case. We also adopt $\vec{\theta_0} = \vec{\beta_\rmS} + \vec{\theta_\rmE}$, which implies that the radius of the unperturbed, fiducial arc is equal to $1.866$ arcsec.

\section{Detectability of the black hole} 
\label{sec:detectability}

Fig.~\ref{lensing_distortion} illustrates how a perturber of mass $M_\rmP = 10^7 \Msun$, located at
$z_\rmp = 0.01$, impacts our fiducial lensing configuration. The dashed line shows the unperturbed arc, while the three solid lines show the perturbed arc, computed by solving for $\delta\vec{\theta}$ using
equation~(\ref{matrixnot}). Different colors correspond to different (angular) impact parameters, as indicated. Note how the distortion becomes more localized (more `kink'-like), and more pronounced, as the black hole comes closer to the unperturbed geodesic. 

Throughout this paper, we shall define the `detectability' of the perturber based purely on this geometrical, kink-like distortion, without taking account of how the perturber magnifies or de-magnifies the light along the arc. We focus exclusively on kink-like distortions that are easily detectable `by eye', by which we mean, without any sophisticated analysis. For this to be the case the magnitude of the distortion needs to be both sufficiently large (compared to the resolution of the data), and sufficiently local or `kink'-like. The magnitude of the distortion can be quantified in terms of the magnitude of the vector $\delta\vec{\theta_1}$ in Fig.~\ref{angles_illustration}, i.e., the magnitude of the radial distortion at the point closest to the black hole. The importance of the second, `locality' criterion is obvious from considering the green curve in Fig.~\ref{lensing_distortion}: although in this case $\vert\delta\vec{\theta_1}\vert$ may be significantly larger than the spatial resolution of the data, $\calR$, the distortion is not particularly `localized'. This owes its origin to the fact that the tangential component of the distortion vector, $\delta\vec{\theta}$, can become much larger than the radial component. If that is the case, the distortion may elude detection because it will be difficult to tell such a distortion apart from the effect of external shear, or a small modification of the (shape of) the main lens. 

In order to assure that the perturbation is sufficiently localized, we therefore quantify the detectability in terms of the {\it difference} in the radial perturbations at two points along the arc. Those two points include the point on the arc closest to the perturbing black hole and another point separated by an angular distance of $f\omega_{\rm{arc}}/2$ along the arc, where $0 < f < 1$ is a free parameter.  As long as this difference in the radial distortion is sufficiently large compared to the spatial resolution of the data, the perturber's presence will be detectable by eye from a localized, kink-like distortion in the razor-thin arc. Hence, our detectability criterion is given by
\begin{equation}
\Delta \theta_\rmr \equiv \delta \theta_\rmr (\vec{\theta_1}) - \delta \theta_\rmr (\vec{\theta_2}) \geq \calR/2\,,
\label{detectcriteria}
\end{equation}
where the indices 1 and 2 refer to the two points along the arc; point 1 is located closest to the perturber where the radial distortion is the largest, and point 2 is offset from point 1 by an angle $f\omega_{\rm{arc}}/2$ along the arc (see Fig.~\ref{lensing_distortion}). Throughout, we adopt $f=0.5$ as our fiducial value. As we demonstrate in \S\ref{sec:single} below, our results only depend very weakly on this choice.

For given redshifts of source, lens and perturber, the distortion quantified in terms of $\Delta \theta_\rmr$ becomes larger with increasing black hole mass, $M_\rmP$, and decreasing impact parameter $b_0$, defined as the distance in the lens plane of the perturber between the perturber and the unperturbed geodesic between source and observer, i.e., 
\begin{align}\label{impactparam}
b_0 \equiv D_{\rmP} \, \Delta \beta_\rmP = \begin{cases}
D_{\rmP} \, \left(|\vec{\theta_0}| - \beta_\rmP\right), & \text{foreground,}\\
D_{\rmP} \, \left(|\vec{\theta^{\prime}_0}| - \beta_\rmP\right), & \text{background,}
\end{cases}
\end{align}
where $|\vec{\theta'_0}| = |\vec{\theta_0}| -\gamma_\rmb \theta_\rmE$, and we have defined the corresponding {\it angular} impact parameter, $\Delta \beta_\rmP$.
\begin{figure}
\centering
\includegraphics[width=0.52\textwidth,height=0.3\textwidth]{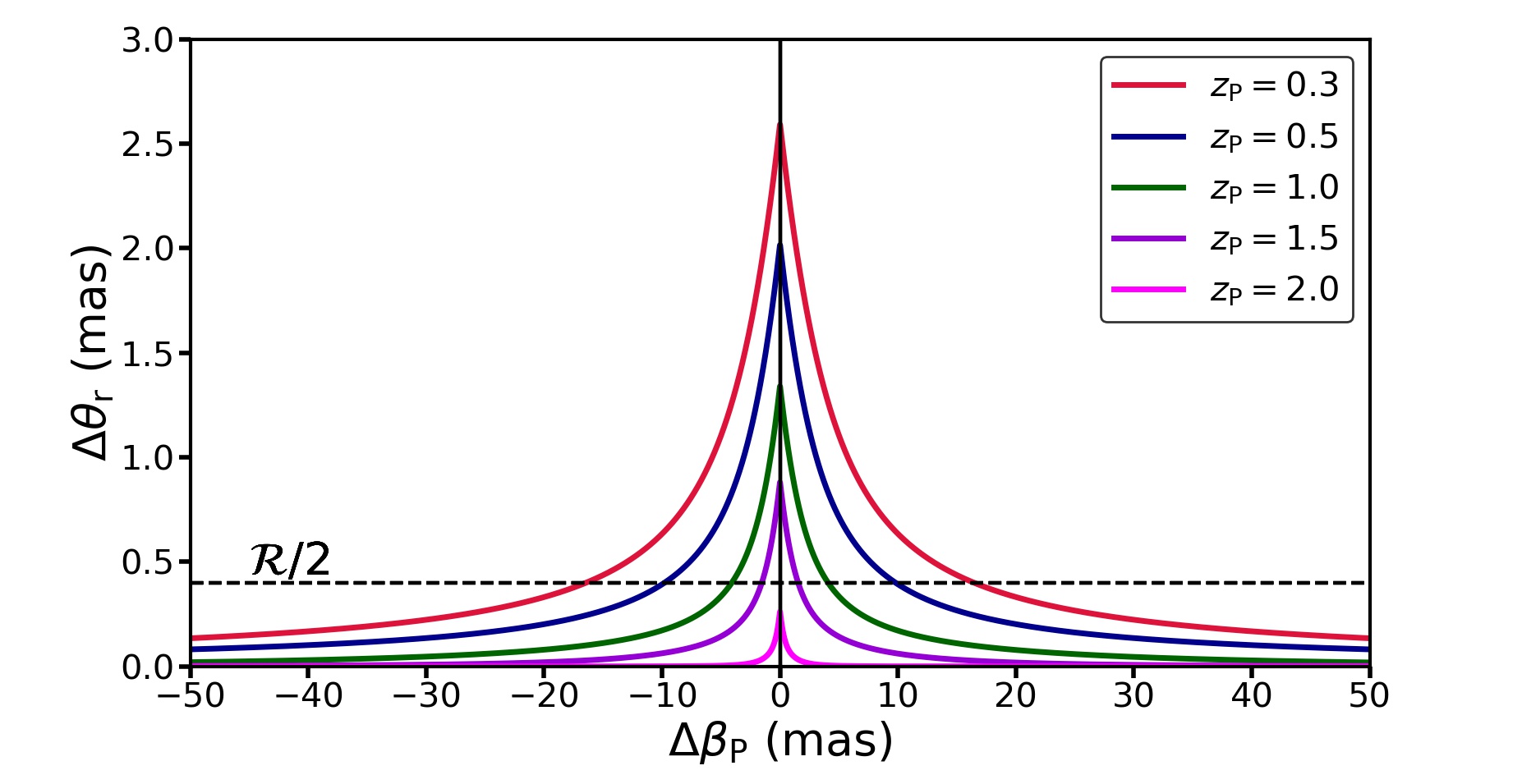}
\caption{\label{fig3} Angular distortion of our fiducial lensing arc, quantified in terms of $\Delta\theta_\rmr$ as a function of angular impact parameter, $\Delta \beta_\rmP$, for a perturbing black hole of mass $M_\rmP=10^6 \Msun$. Results are shown for five different redshifts of the perturber (red and blue in the foreground and the rest in the background), as indicated. The dotted, horizontal line indicates the detection threshold, $\calR/2$, where we adopt our fiducial, spatial resolution of $\calR = 0.8$ mas. The perturbation is deemed detectable whenever the curve is above this threshold, i.e., when
$\Delta\theta_\rmr > \calR/2$). This occurs for a finite range of angular impact parameters, $|\Delta \beta_\rmP| < \Delta\beta_{\rm P,max}$. Note that $\Delta\beta_{\rm P,max}$ decreases with increasing redshift of the perturber and that the maximum distortion (which occurs for zero impact parameter and is equal to the Einstein angle of the perturber) falls below the detection limit beyond a certain redshift.}
\end{figure}

Fig.~\ref{fig3} plots the distortion, $\Delta \theta_\rmr$, for our fiducial lensing configuration (\S\ref{sec:fiducial}) as a function of the angular impact parameter, $\Delta \beta_\rmP$, computed using equations~(\ref{matrixnot}) and (\ref{radtandeltheta}) for a perturber mass of $M_\rmP = 10^6 \Msun$. The different curves correspond to different redshifts for the perturber (two in the foreground and three in the background), as indicated, while the dashed, horizontal line corresponds to a fiducial spatial resolution of $\calR/2 = 0.4$ mas. Placing the perturbing black hole at a smaller redshift results in a larger distortion, and hence in a wider range of the angular impact parameter for which $\Delta \theta_\rmr > \calR/2$. This is because the Einstein radius of the perturber decreases with increasing redshift.

\begin{figure}
\centering
\includegraphics[width=0.52\textwidth,height=0.3\textwidth]{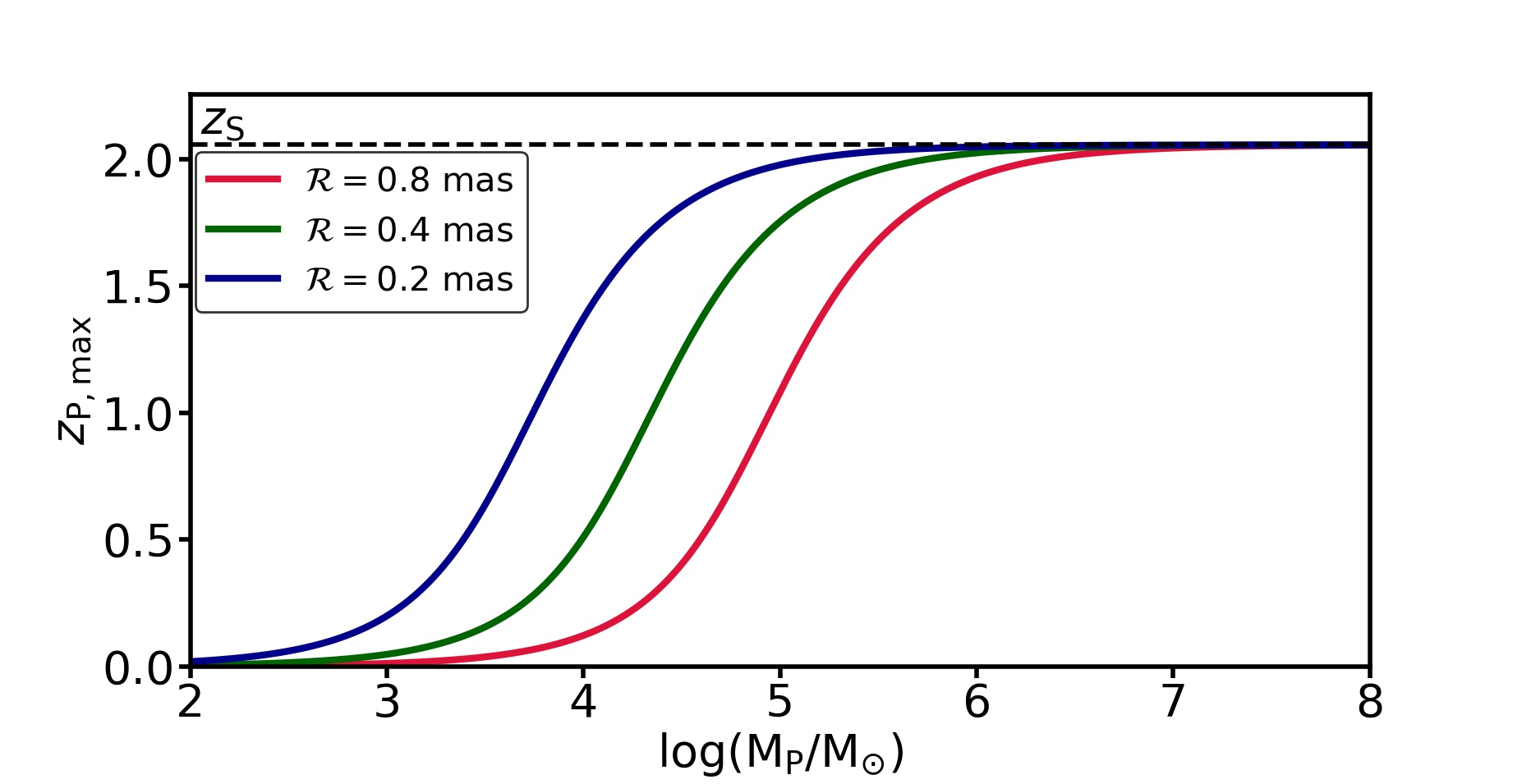}
\caption{\label{fig4} Maximum redshift out to which kink-like distortions of our fiducial arc due to a perturbing black hole are detectable as a function of the black hole mass $M_\rmP$. Different colors correspond to different spatial resolutions, as indicated.  The maximum redshift for detectability increases with black hole mass and tends towards the source redshift for sufficiently massive black holes. Also note that it increases for higher spatial resolution.}
\end{figure}

At given redshifts for source, lens and perturber, the condition for detectability translates into a required range for the angular impact parameter,
\begin{equation}\label{deltabetarange}
\Delta \beta_\rmP \leq \Delta \beta_{\rm P,max}\,, 
\end{equation}
where $\beta_{\rm P,max}$ corresponds to the angular impact parameter for which $\Delta \theta_\rmr = \calR/2$. Note that a sufficiently small perturber may never satisfy the detectability condition stated in equation (\ref{detectcriteria}), even for zero impact parameter. In fact, to each perturber redshift, $0 \leq z_\rmP \leq z_\rmS$, corresponds a minimum detectable perturber mass, $M_{\rm P, min}$, or, equivalently, with each perturber mass $M_\rmP$, corresponds a maximum redshift, $0 \leq z_{\rm P, max} \leq z_\rmS$, out to which such a perturber can be detected. Beyond that redshift even the maximum value of $\Delta \theta_\rmr$, which occurs for $\Delta \beta_\rmP=0$ and is equal to $\theta_{\rm EP}$, falls below the detection limit $\calR/2$. Therefore the maximum redshift of detectability, $z_{\rm{P,max}}$, can be obtained as the root for $z_\rmP$ of $\theta_{\rm{EP}}(z_\rmP) = \calR/2$. The red line in Fig.~\ref{fig4} plots $z_{\rm P,max}$ as a function of $M_\rmP$ for our fiducial lens configuration. It increases from close to zero for $M_\rmP < 10^3 \Msun$ to roughly $z_\rmS$ for $M_\rmP > 10^7 \Msun$. Hence, free-floating black holes less massive than $\sim 10^3\Msun$ are never detectable (according to our detectability criterion), while those with $M_\rmP \gta 10^7\Msun$ are detectable at all redshifts between observer and source. The green and blue curves show the behaviour for $\calR=0.4$ and $0.2$ mas respectively. As expected, the maximum redshift of detectability increases with better spatial resolution (smaller $\calR$).

\begin{figure}
\centering
\includegraphics[width=0.52\textwidth,height=0.3\textwidth]{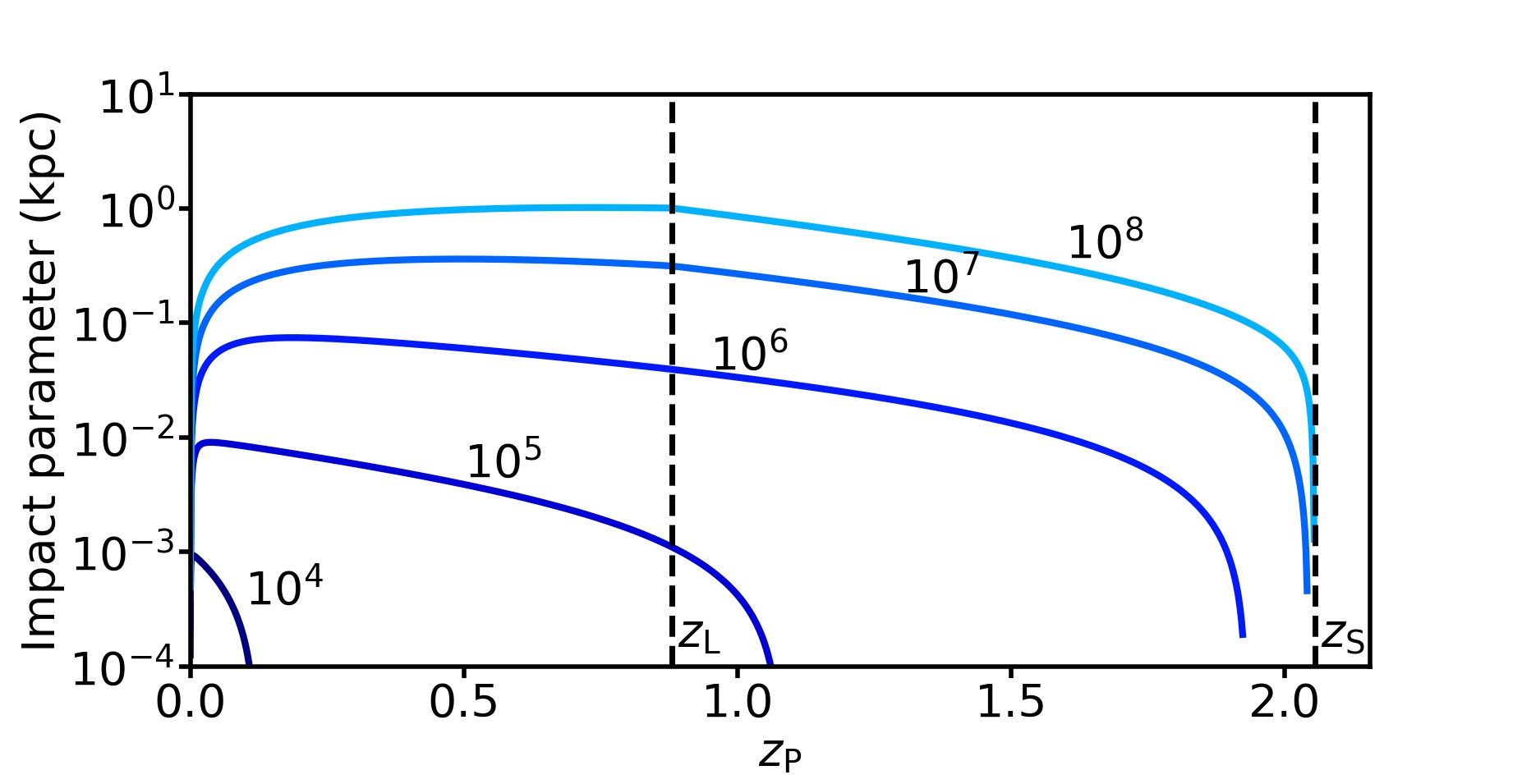}
\caption{\label{fig5} Maximum impact parameter for which a perturbing black hole causes a detectable distortion of our fiducial arc. Results are shown as a function of the redshift of the perturber, $z_\rmP$, and for 5 different perturber masses, as indicated. The maximum impact parameter increases with perturber mass but initially increases and then decreases with perturber redshift. Note that the maximum redshift out to which a perturbing black hole can be detected increases with black hole mass, and approaches the source redshift for $M_\rmP \gta 10^7\Msun$.}
\end{figure}

Fig.~\ref{fig5} plots the maximum impact parameter, $b_{0,\rm max} = D_{\rmP}\Delta\beta_{\rm P, max}$, as a function of the redshift of the perturbing black hole, $z_\rmP$, for our fiducial lensing configuration. Different curves correspond to different black hole masses, as indicated. Note how perturbing black holes with mass $M_\rmP \lta 10^7 \Msun$ can only be detected out to a redshift $z_{\rm P, max} < z_\rmS$ (cf. Fig.~\ref{fig4}). One can also see from Fig.~\ref{fig5} that $b_{0,\rm max}$ increases with higher perturber mass.

\begin{figure*}
\centering
\includegraphics[width=1\textwidth,height=0.3\textwidth]{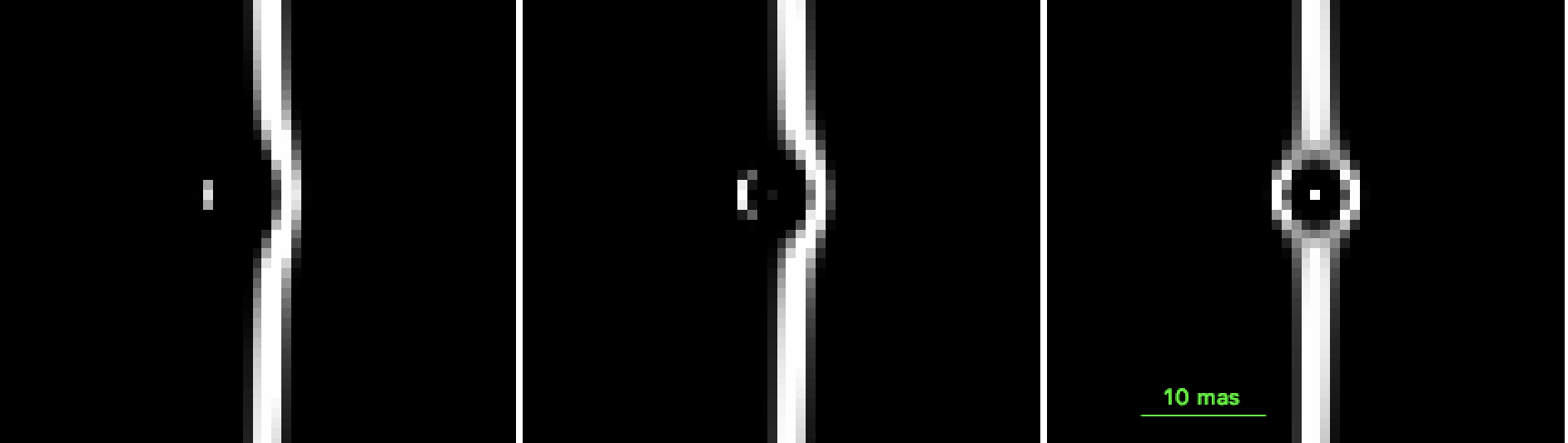}
\caption{Mock (noiseless) images constructed using {\tt GRAVLENS} \citep{Keeton.01} of the distortions caused in a razor-thin lensing arc due to a perturbing BH of mass $M_\rmP = 10^6 \Msun$ at $z_\rmP=0.5$ along the line of sight. The images have been obtained by boxcar filtering with a FWHM of 0.8 mas. An arc $\sim 270$ mas long is formed due to the gravitational lensing of a source at $z_\rmS = 2.056$ by a singular isothermal sphere with a mass of $\sim 10^{12} \Msun$ within the Einstein radius ($\approx 2"$), and located at a redshift $z_\rmL = 0.881$. The source is assumed to have a Sersic profile and is aligned at an angle of $36$ mas from the baseline. The radius of the unperturbed arc is $\approx 2.036"$ which is equivalent to 2545 pixels (each pixel $\equiv 0.8$ mas) in this image. The different columns from left-to-right correspond to different angular impact parameters ($\Delta \beta_\rmP$) of $-3.2$, $-1.6$ and $0$ mas with respect to the unperturbed arc. Note how the BH causes a kink-like distortion in the arc, which is easily detectable `by eye', given the resolution of the images, and in the absence of noise.}
\label{illustration}
\end{figure*}

As already eluded to above, our detectability criterion is based solely on the geometric distortion $\delta\vec{\theta}$, without recourse to the magnification caused by the perturber. In order to have an idea as to the potential impact of magnification, we have used {\tt GRAVLENS} \citep{Keeton.01} to make mock images of our fiducial razor-thin arc, being perturbed by a black hole of mass $M_\rmP = 10^6 \Msun$ at a redshift of $z_\rmP = 0.5$. The resulting images, for a spatial resolution of $0.8$~mas, are shown in  Fig.~\ref{illustration}. From left to right, the different panels correspond to angular impact parameters of $\Delta\beta_\rmP = -3.2$, $-1.6$, and $0$~mas. The left and middle panels reveal local, kink-like distortions similar to what is shown in Fig.~\ref{lensing_distortion}. In addition, the perturber creates a small, second arclet opposite to the kink, which is a feature that we ignore throughout this paper. When the perturber exactly aligns with the arc, as in the right-hand panel, a small Einstein ring is visible. Upon close inspection of these images, it is evident that the perturber causes some magnification/demagnification of parts of the arc close to the distortion, but overall it is clear that the main characteristic of the distortion is its kink-like geometry, not the corresponding (de)magnification.  This justifies the use of our detectability criterion~(\ref{detectcriteria}). In fact, by focusing only on the geometric distortion, our constraints will be conservative, i.e. a sophisticated analysis of the surface brightness variations along the arc, similar to the analyses of strong-lensing distortions by \citet{Vegetti.etal.14}, \citet{Hezaveh.etal.16b}, or \citet{Birrer.etal.17}, might allow the detection of even lower-mass perturbers due to their localized (de)magnification of the arc.

%

We point out that throughout we assume the main lens to be spherically symmetric. In general, the shape of the kink might change, and the kink itself might be more difficult to identify `by eye', if the main lens is strongly elliptical or if external shear is present. This could potentially make the approach presented here less effective. However, in general such extreme configurations would also disrupt the arc, splitting it in multiple images, and thus leading to a different type of lensing configuration. Hence, given that this is only a fairly crude analysis, based on distortions that are easily identifiable `by eye', and given that only a handful of systems have been observed yet at milli-arcsecond resolution, we consider the influence of ellipticity and shear to be higher-order effects that only warrant careful consideration when the quantity and quality of the data improve.

Finally, we emphasize that the localized kink-like distortions considered here must arise from extremely compact objects, such as the black holes. They cannot be caused by perturbations due to the (far more abundant) dark matter haloes along the line of sight. Although dark matter haloes (or subhaloes) can perturb gravitational lensing arcs \citep[see e.g.,][]{Koopmans.05, More.etal.09, Vegetti.Koopmans.09, Vegetti.etal.10, Vegetti.etal.12, Hezaveh.etal.16b, Birrer.etal.17}, we demonstrate in Appendix~\ref{sec:subhalo} that such perturbations are not sufficiently localized to be detectable according to our criterion (\ref{detectcriteria}). 

\section{Towards constraints on the comoving number density of free-floating black holes}
\label{sec:Veff}

Our aim in this paper is to translate the presence of kink-like distortions along razor-thin lensing arcs, or the absence thereof, into constraints on the number density of black holes; either primordial ones or wandering black holes that were formed as a by-product of galaxy formation (see \S\ref{sec:intro}). In the previous two sections, we have shown how to compute the angular extent of such a kink-like distortion, $\Delta \theta_\rmr$, for a given lensing configuration and a given black hole mass, and we have shown that the criterion for detectability translates into a constraint on the maximum (angular) impact parameter of the black hole with respect to the unperturbed geodesic from the source to the observer. We now show how a given number of kink-like distortions, $\Ndis$, for a lensing arc of length $\arc$, translates into a constraint on the comoving number density of black hole perturbers, $\nBH$. For the sake of simplicity we ignore potential evolution in $n_{\rm BH}$ with redshift, but that is easily accounted for. 

If we assume, for simplicity, that the number of free-floating black holes of mass $\MBH$ in a given comoving volume, $V$, follows a Poisson distribution with mean $\averNBH = \nBH(\MBH) \, V$, then the detection of $\Ndis$ distortions along an arc specified by $z_\rmS$, $z_\rmL$, $\arc$, and $\calR$ implies the following 95 percent confidence interval on $\nBH(\MBH)$:
\begin{equation}\label{nconstraint}
\frac{\averNBH_{\rm lower}}{V_{\rm eff}(\MBH)} \leq \nBH(\MBH) \leq \frac{\averNBH_{\rm upper}}{V_{\rm eff}(\MBH)}\,.
\end{equation}
Here, $V_{\rm eff}$ is the effective volume inside of which a perturbing black hole of mass $\MBH$ can be detected through its kink-like distortion of the arc, which is given by,
\begin{equation}\label{comovingvol}
V_{\rm eff}(M_\rmP) = \int_{0}^{z_\rmS} \Omega (z) \, \frac{\rmd^2 V}{\rmd\Omega \, \rmd z} \, \rmd z\,.
\end{equation}
Here, $\rmd^2 V/(\rmd\Omega \, \rmd z)$ is the comoving volume element at redshift $z$ corresponding to a solid angle $\rmd \Omega$ and a redshift depth $\rmd z$, and the solid angle $\Omega(z)$, in the small angle approximation, is given by 
\begin{equation}\label{Omega}
\Omega(z) \approx 2 \arc\,\Delta \beta_{\rm P, max}(z)\,.
\end{equation}
The factor of $2$ accounts for the fact that the perturber may reside on either side of the arc, and the angular impact parameter $\Delta \beta_{\rm P, max}(z)$ follows from our detection criterion (equation~\ref{detectcriteria}) and the double lens equations presented in Sections~\ref{sec:foreground} and \ref{sec:background}. Note that $\Omega(z) = 0$ for $z > z_{\rm P,max}$. The total effective volume $V_{\rm eff}$ can be split into the foreground and background volumes, which we compute separately and then add up.

The $1-\calC$ confidence interval on $\averNBH$ is given by
\begin{equation}\label{lam_lower}
\averNBH_{\rm lower} \equiv \frac{1}{2} \,\chi^2\left(\calC/2\,, 2\Ndis\right) \,,
\end{equation}
and
\begin{equation}\label{lam_upper}
\averNBH_{\rm upper} \equiv \frac{1}{2} \,\chi^2\left(1-\calC/2\,, 2\Ndis+2\right)\,,
\end{equation}
where $\chi^2(p;n)$ is the quantile function of the $\chi^2$-distribution with $n$ degrees of freedom \citep[][]{Garwood.36}. Values for $\averNBH_{\rm lower}$ and $\averNBH_{\rm upper}$ have to be computed numerically or taken from look-up tables. For easy reference, Table~\ref{tab:CL} lists  $\averNBH_{\rm lower}$ and $\averNBH_{\rm upper}$ for $\Ndis = 0,1,2,\ldots,5$ and for confidence levels of $\calC = 0.32$, $0.05$, and $0.01$. Note that in the absence of any detections ($\Ndis=0$), we only obtain an upper limit on the number density of black holes; at 95 percent (99 percent) confidence, a null-detection implies that $\nBH(\MBH) \, V_{\rm eff}(\MBH) < 3.69$ ($5.30$). 

The above applies to the constraints that result from a single arc. When multiple arcs have been observed, the constraint on $\nBH(\MBH)$ simply follows from adding the effective volumes and distortions of the individual arcs, that is, from using equations~(\ref{nconstraint})-(\ref{lam_upper}) with $V_{\rm eff} \rightarrow V_{\rm eff, tot} = \sum_{i=1}^{N_{\rm arc}} V_{{\rm eff},i}$ and $\Ndis \rightarrow N_{\rm dis,tot} = \sum_{i=1}^{N_{\rm arc}} N_{{\rm dis},i}$.
 
\begin{figure*}
\includegraphics[width=1.1\textwidth]{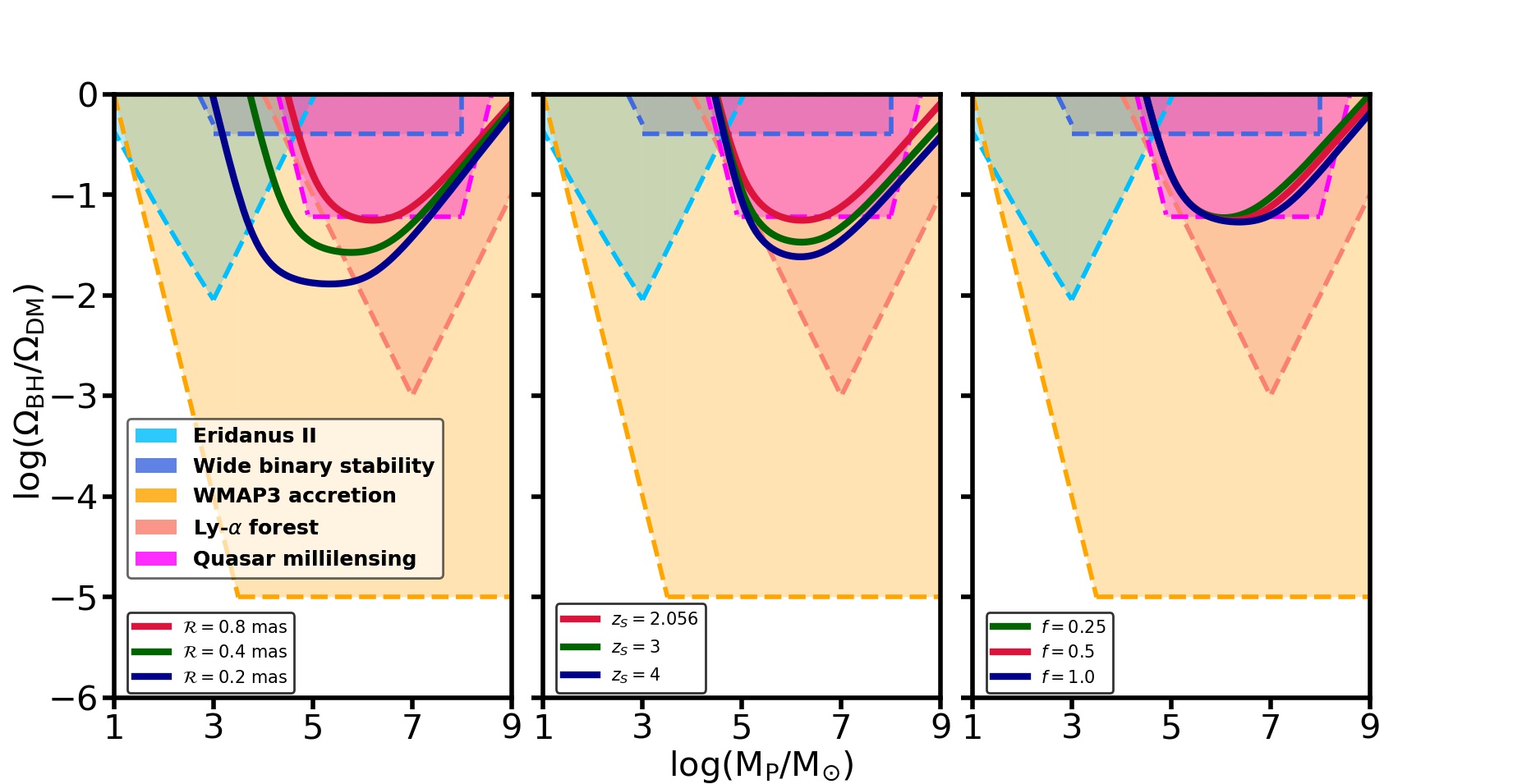}
\caption{\label{fig6} Single null detection: maximum constraint on the ratio of the mean comoving mass density of free-floating black holes (considered as primordial black holes) to the dark matter comoving density expressed as $\Omega_{\rm {BH}}/\Omega_{\rm {DM}}$ vs black hole mass $M_{\rmP}$ in units of $\Msun$ for different values of $\calR$, $z_\rmS$ and $f$ ($M_{\rmL}=10^{12} \Msun$, $\omega_{\rm arc}=270$ mas, $z_\rmL=0.881$) at $95\%$ confidence level. The constraints in general are the tightest at some intermediate black hole mass the value of which depends on the spatial resolution $\calR$ with which the arc is probed. Also shown are the previously obtained constraints (shaded regions are ruled out) on $\Omega_{\rm {BH}}/\Omega_{\rm {DM}}$ for PBHs. From the left panel it can be seen that the constraints get tighter and the characteristic black hole mass (turnover mass) gets reduced for higher spatial resolution. In the near future, with the discovery of more finely resolved lensing arcs, it will be possible to put constraints on the unconstrained regions of the plot, i.e. in the mass range $M_{\rm P}\lesssim 10^3 \Msun$. Also shown are existing constraints from quasar milli-lensing \citep{Wilkinson.etal.01}, the survival of a star cluster in Eridanus II \citep[][]{Brandt.etal.16}, wide binary stability \citep{Quinn.Smith.09}, the impact of Poisson noise in the matter power spectrum on the Ly-$\alpha$ forest \citep{Afshordi.etal.03}, and WMAP3 constraints on accretion onto black holes prior to recombination \citep{Ricotti.etal.08}.}
\end{figure*}

\section{Results}
\label{sec:massdensityconstraint}

In this section, we discuss various constraints on the number and mass densities of free-floating black holes that one may realistically achieve with observations of our fiducial razor-thin arc. 

\subsection{Constraints for single-mass species}
\label{sec:single}

We start by considering the case in which all free-floating black holes have the same mass, $\MBH$. Under the assumption of no detected distortions (i.e., $\Ndis = 0$), we obtain an upper limit on the mass density of such black holes, $\nBH$, which we express in terms of $\Omega_{\rm BH}/\Omega_{\rm DM,0}$. Here $\Omega_{\rm DM,0}$ is the assumed dark matter matter density in units of the critical density, for which we adopt $\Omega_{\rm DM,0} = 0.259$ \citep{Planck.14}, and $\Omega_{\rm BH} = \MBH \, \nBH / \rho_{\rm crit}$.
\begin{table}
\centering
\caption{Two-sided confidence intervals $(\averNBH_{\rm lower},\averNBH_{\rm upper})$ for the mean of a Poisson distribution given $\Ndis$ detections. We list results for $\Ndis=0,1,\ldots,5$ and for confidence levels of 68 percent ($\averNBH_{0.160},\averNBH_{0.840}$), 95 percent ($\averNBH_{0.025},\averNBH_{0.975}$), and 99 percent ($\averNBH_{0.005},\averNBH_{0.995}$).}
\begin{tabular}{l|l|l|l|r|r|r}
\hline
 & \multicolumn{3}{|c|}{$\averNBH_{\rm lower}$} & \multicolumn{3}{|c|}{$\averNBH_{\rm upper}$} \\
$\Ndis$ & $\averNBH_{0.005}$ & $\averNBH_{0.025}$ & $\averNBH_{0.160}$ & $\averNBH_{0.840}$ & $\averNBH_{0.975}$ & $\averNBH_{0.995}$\\
\hline\hline
0 & $0.00$ & $0.00$ & $0.00$ & $1.83$ & $3.69$ & $5.30$ \\
1 & $0.01$ & $0.03$ & $0.17$ & $3.29$ & $5.57$ & $7.43$ \\
2 & $0.10$ & $0.24$ & $0.71$ & $4.62$ & $7.22$ & $9.27$ \\
3 & $0.34$ & $0.62$ & $1.37$ & $5.90$ & $8.77$ & $10.98$ \\
4 & $0.67$ & $1.09$ & $2.09$ & $7.15$ & $10.24$ & $12.59$ \\
5 & $1.08$ & $1.62$ & $2.85$ & $8.37$ & $11.67$ & $14.15$ \\
\hline
\end{tabular}
\label{tab:CL}
\end{table}

The solid lines in Fig. \ref{fig6} show the constraints on $\Omega_{\rm {BH}}/\Omega_{\rm {DM}}$ that we obtain as a function of the black hole mass $M_{\rmP}$ for different resolutions (left-hand panel), different source redshifts (middle panel), and different values of the parameter, $f$ (right-hand panel). In each panel, the red curve corresponds to our fiducial configuration. The constraints on $\Omega_{\rm {BH}}/\Omega_{\rm {DM}}$ are most stringent for intermediate black hole mass with $M_\rmP \sim 10
^6 \Msun$. The constraints improve with higher spatial resolution (i.e., smaller $\calR$), especially at the low mass end, with increasing source redshift, $z_\rmS$, and depend only weakly on the parameter $f$
that regulates the angular distance along the arc at which we compare the angular distortion, $\delta\theta_\rmr$. The trend with $z_\rmS$ is easy to understand from the fact that the effective volume increases with increasing $z_\rmS$. Similarly, decreasing $\calR$ and increasing $f$ make our detectability criterion~(\ref{detectcriteria}) less strict, thereby resulting in a stronger constraint on $\Omega_{\rm {BH}}/\Omega_{\rm {DM}}$. Note that the constraints on $\Omega_{\rm {BH}}/\Omega_{\rm {DM}}$ get tighter with increase in the arc length probed, $\arc$, i.e. as more razor-thin arcs are observed.

In Appendix \ref{sec:scale}, we derive how our constraints scale with the mass of the perturber, $M_\rmP$, and with the spatial resolution, $\calR$, by analytically solving the perturbative form of the double lens equation~(\ref{matrixnot}) in the small and large mass limits. In particular, we show that for massive perturbers, which have a mass well in excess of a characteristic mass scale
\begin{equation}
M_0 \equiv \frac{\calR^2 c^2}{16\, G} \, D_{\rmS} \simeq 3 \times 10^4 \Msun
\, \left(\frac{\calR}{\rm mas}\right)^2 \, \left(\frac{D_{\rmS}}{\rm Gpc}\right)\,,
\end{equation}
the effective volume (equation~[\ref{comovingvol}]) scales as
\begin{equation}
V_{\rm eff} \propto {\left(\arc\right)}^{\frac{5}{3}} \, {\left(\frac{M_\rmP}{\calR}\right)}^{\frac{1}{3}}\,.
\end{equation}
Since $\Omega_{\rm BH} \propto \rho_{\rm BH} \propto M_{\rm BH}/V_{\rm eff}$, the
constraint on $\Omega_{\rm {BH}}/\Omega_{\rm {DM}}$ for a given number of detections, $\Ndis$, scales as

\begin{equation}
\Omega_{\rm {BH}}/\Omega_{\rm {DM}} \propto {M_\rmP}^{\frac{2}{3}} {\calR}^{\frac{1}{3}} \frac{\Ndis}{{\left(\arc\right)}^{\frac{5}{3}}}\,.
\end{equation}
For perturbers with a mass $M_\rmP \ll M_0$, the effective volume scales as,
\begin{equation}
V_{\rm eff} \propto \arc \, \frac{M^3_\rmP}{\calR^5}\,,
\end{equation}
which implies that
\begin{equation}
\Omega_{\rm {BH}}/\Omega_{\rm {DM}} \propto \frac{1}{\arc} \, \frac{\calR^5}{M^2_\rmP} \, \Ndis\,.
\end{equation}
Hence, the constraint on $\Omega_{\rm {BH}}/\Omega_{\rm {DM}}$ becomes much weaker for lower mass black holes. Note, though, that the constraining power depends extremely strongly on the spatial resolving power in this limit. Increasing the spatial resolution by a factor of 2  will improve the constraints on $\Omega_{\rm {BH}}/\Omega_{\rm {DM}}$ (for $M_\rmP \ll M_0$) by a factor of 32. In addition, the characteristic turnover mass $M_{\rm TO}$ (see equation [\ref{turnovermass}] of Appendix \ref{sec:scale}) that scales as ${\calR}^{7/4}$ reduces by a factor of $\approx 3.4$ that further improves the constraining power. 

It is informative to compare these constraints from a null-detection along our fiducial arc to existing constraints. The shaded regions in Fig.~\ref{fig6} show such constraints\footnote{To not clutter Fig.~\ref{fig6}, the constraints shown are not exhaustive; see \citet{Carr.Silk.18} for a few additional constraints in the mass range shown.} from quasar milli-lensing \citep{Wilkinson.etal.01}, the survival of a star cluster in Eridanus II \citep[][]{Brandt.etal.16}, wide binary stability against tidal disruption by black holes \citep{Quinn.Smith.09}, generation of large-scale structure through Poisson fluctuations and its imprint on Ly $\alpha$ clouds \citep{Afshordi.etal.03}, and WMAP constraints on the accretion effects of PBHs on the CMB \citep{Ricotti.etal.08}. The WMAP and Ly $\alpha$ constraints apply exclusively to PBHs, whereas the other constraints from milli-lensing and dynamical effects apply to all black holes independent of their formation epoch. 

Clearly, the constraints from a potential null-detection along our fiducial razor-thin arc are not particularly competitive. For the fiducial redshifts of the source and the lens, and for a fiducial resolution of $\calR=0.8$~mas, the constraints are comparable to those from the quasar milli-lensing constraints of \citet{Wilkinson.etal.01}, which derive from the absence of multiple images (at mas resolution) among a sample of 300 compact radio sources. These constraints, though, are much weaker than the constraints (on PBHs) that derive from WMAP or Ly-$\alpha$ data. In order for distortions of razor-thin arcs to yield constraints that are competitive with these data we require a sample of hundreds to thousands of razor-thin arcs, preferentially at high spatial resolution, and with high-redshift sources.

\subsection{Constraining the mass function of wandering black holes}
\label{sec:massfunc}

\begin{figure}
\centering
\includegraphics[width=0.3\textwidth,height=0.45\textwidth,angle=270]{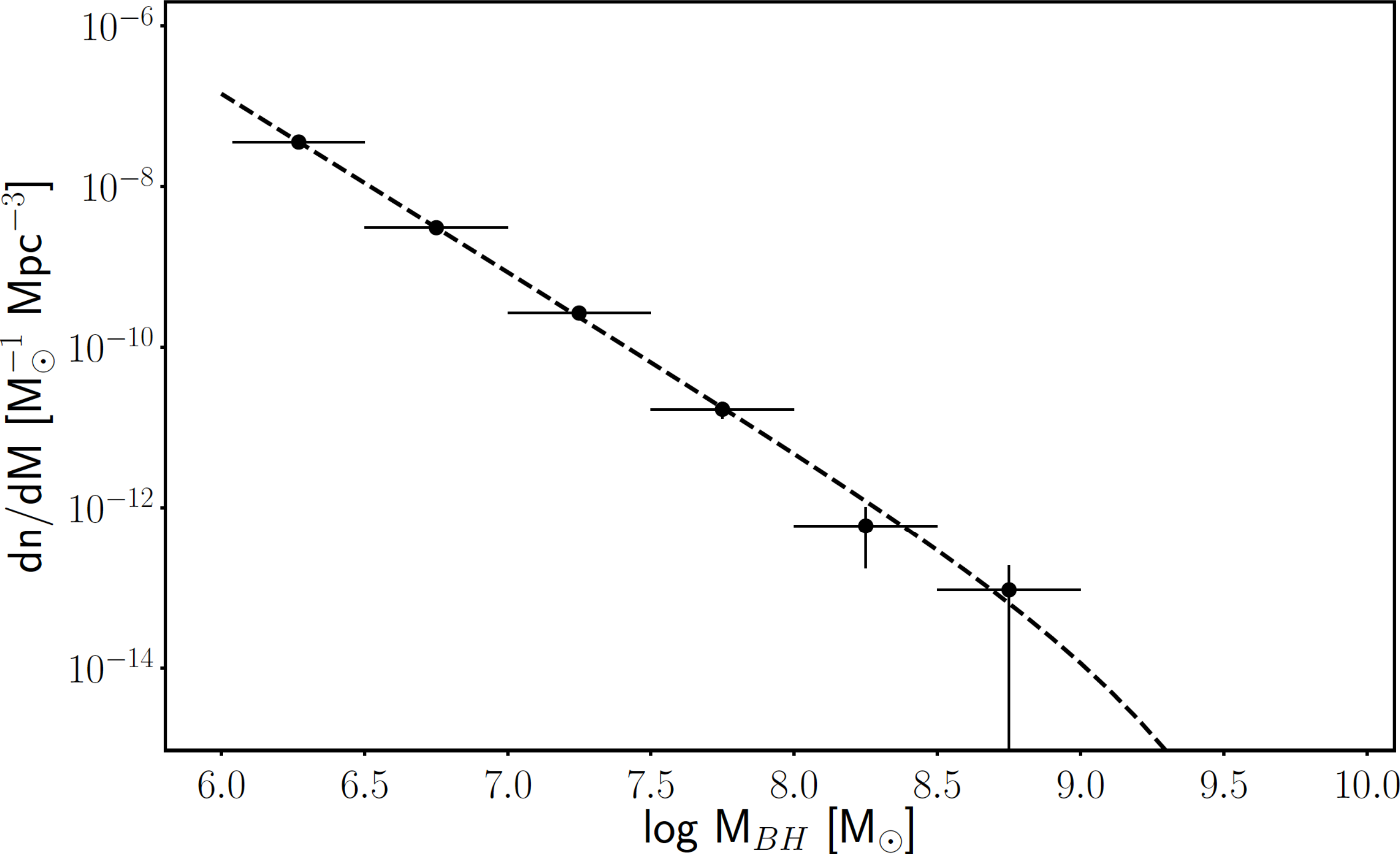}
\caption{\label{fig:BHmassfunc} Mass function of wandering black holes ($M_{\rm BH} \gtrsim 10^6 \Msun$) computed from the cosmological simulation {\sc Romulus25} and the best fit Schechter function with $M^*_{\rm BH}=10^9 \Msun$ and the best fit values of $n^*_{\rm BH}=3.11\times 10^{-5}~{\rm Mpc}^{-3}$ and $\alpha = -2.22$.}
\end{figure}

\begin{figure*}
\centering
\includegraphics[width=1\textwidth,height=0.5\textwidth]{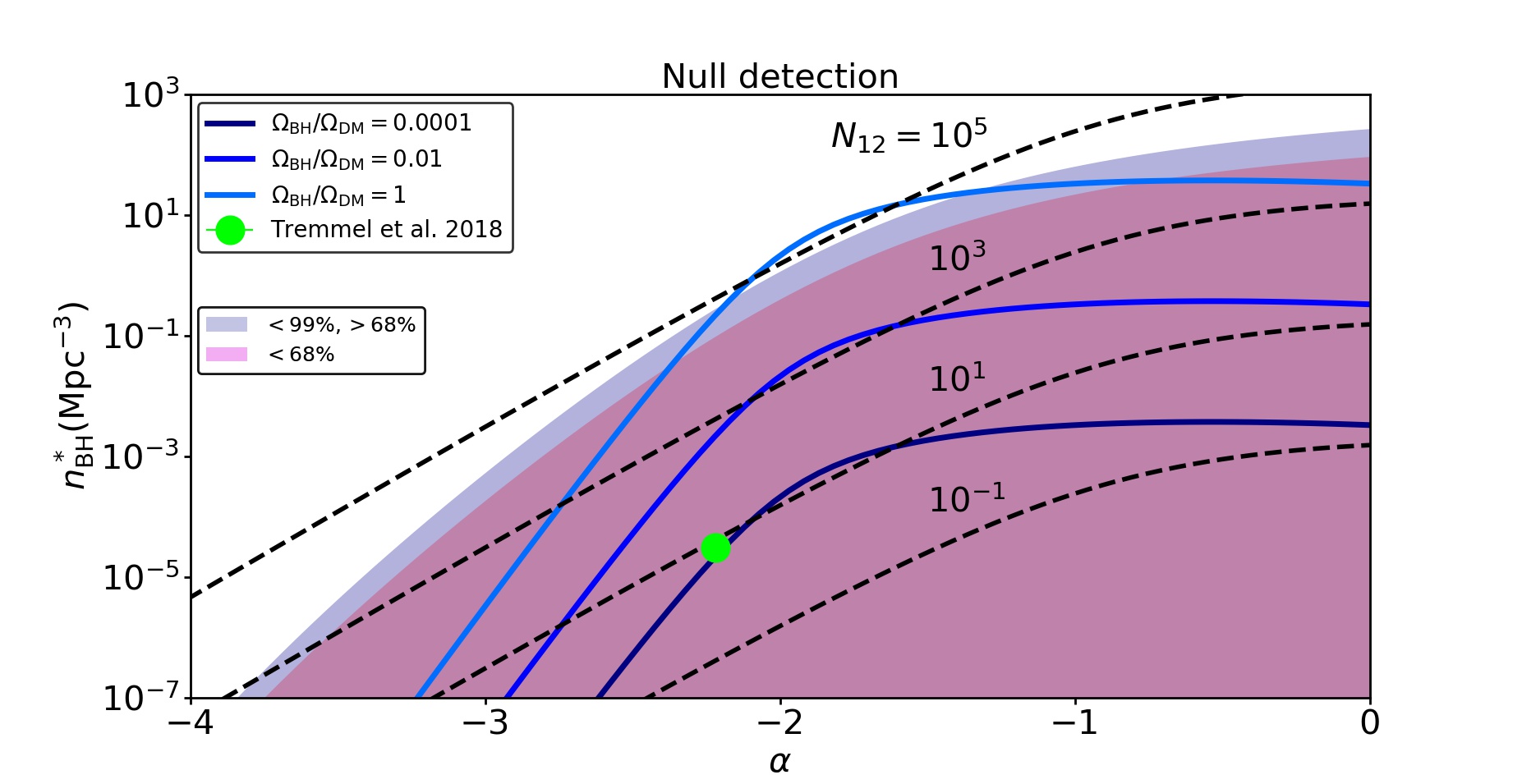}
\caption{Single null detection: contours of $n^*_{\rm BH}$ vs $\alpha$ for different values of the occupation number $N_{12}$ for wandering black holes (dashed lines) and for different values of $\Omega_{\rm {BH}}/\Omega_{\rm {DM}}$ (solid lines). The shaded region represents the allowed values of $n^*_{\rm BH}-\alpha$ for a null detection of lensing distortion events, given the fiducial configuration ($z_\rmS=2.056$, $z_{\rmL}=0.881$, $\calR=0.8$ mas, $M_\rmL(\theta_\rmE) = 10^{12} \Msun$). The two different shades from top to bottom indicate the 99 and 68 percent confidence levels. The green dot represents the values of $n^*_{\rm BH}$ and $\alpha$ obtained by \citet{Tremmel.etal.18} in their cosmological simulation {\sc Romulus25}. }
\label{fig7a}
\end{figure*}

We now consider the more realistic case, in which the free-floating black holes are characterized by a mass function. In the case of primordial black holes, it is unclear what the mass function will be; depending on the exact formation mechanism and epoch, the mass function can be either extended or very narrow \citep[see discussion in][]{Carr.etal.16}. In the case of wandering black holes, though, we may use the notion that they are a by-product of galaxy formation to argue that they should (roughly) follow a Schechter mass function,
\begin{equation}\label{BHmassfunc}
\frac{\rmd n}{\rmd \MBH} \, \rmd \MBH = n^*_{\rm BH} \, \left(\frac{\MBH}{\Mchar}\right)^{\alpha} \, \exp\left(-\frac{\MBH}{\Mchar}\right) \, \frac{\rmd \MBH}{\Mchar}\,.
\end{equation}
After all, the galaxy stellar mass function is well described by a Schechter function, and the masses of SMBHs at the centres of galaxies are tightly correlated with bulge mass \citep[][]{Marconi.Hunt.03, Haring.Rix.04}. Using that the most massive galaxies have SMBHs of mass $M_{\rm BH} \sim 10^9 \Msun$, we adopt a characteristic black hole mass of $\Mchar = 10^9 \Msun$. Above this mass scale the abundance of wandering black holes is exponentially suppressed, while the mass function follows a simple power-law, with index $\alpha$, for $M \ll \Mchar$. The characteristic, comoving number density, $n^*_{\rm BH}$ sets the overall normalization. 

To demonstrate that the Schechter function is adequate, Fig.~\ref{fig:BHmassfunc} plots the mass function of wandering black holes in the {\sc Romulus25} simulation (solid dots). The dashed line is the best-fit Schechter mass function with $\Mchar = 10^9 \Msun$, which accurately fits that data. The resulting best-fit slope and normalization are $\alpha = -2.22$ and $n^*_{\rm BH} = 3.11 \times 10^{-5} \Mpc^{-3}$. Note that the slope is steeper than $-2$, which implies that the mass density of wandering black holes in the {\sc Romulus25} simulation is dominated by the least massive ones. However, {\sc Romulus25} adopts a black hole seed mass of $10^6 \Msun$, and the results shown in Fig.~\ref{fig:BHmassfunc} are likely affected by this choice. We therefore caution that the predicted slope is likely to change with a simulation of higher mass resolution and/or lower black hole seed mass. 

The mass density of wandering black holes that follow a Schechter mass function is
\begin{equation}\label{rhobh}
\begin{split}
\rho_{\rm BH} & = \int_{M_{\rm min}}^{\infty} \frac{\rmd n}{\rmd \MBH} \, \MBH \, \rmd \MBH \\
 & = n^*_{\rm BH} \, \Mchar \, \Gamma(\alpha+2,M_{\rm min}/\Mchar)\,,
\end{split}
\end{equation}
with $\Gamma(a,x)$ the upper incomplete Gamma function. Here, we have introduced a lower mass limit, $M_{\rm min}$, in order to avoid singularities for integer values of $\alpha \leq -2$. Throughout we adopt $M_{\rm min} = 10^2 \Msun$, which roughly corresponds to the mass scale below which perturbing black holes are no longer detectable (see Fig. \ref{fig4}). Similarly, the expectation value for the number of distortions along a given lensing arc is given by
\begin{equation}\label{Nbh}
\begin{split}
\lambda_{\rm dis} & = \int_0^{\infty} V_{\rm eff}(\MBH) \,  \, \frac{\rmd n}{\rmd \MBH} \, \rmd \MBH \\
& = n^*_{\rm BH} \, \int_0^{\infty}  V_{\rm eff}(x \Mchar)\, x^{\alpha} \, e^{-x} \,,
\end{split}
\end{equation}
where $V_{\rm eff}(\MBH)$ is given by equation~(\ref{comovingvol}). Note that, since $V_{\rm eff} \rightarrow 0$ for $\MBH \lta 10^3\Msun$, here we do not need to cut off the mass integral below $M_{\rm min}$. 

Using these expressions, we can transform the confidence interval $(\averNBH_{\rm lower}, \averNBH_{\rm upper})$ corresponding to a given number of detected distortions, $\Ndis$, into corresponding constraints on $n^*_{\rm {BH}}$ and $\alpha$. In particular, for a null-detection, we have that $\lambda_{\rm dis} < 1.83$ and $5.30$ at 68 and 99~percent confidence levels, respectively. Using equation~(\ref{rhobh}), these constraints can then be transformed into constraints on $\Omega_{\rm BH}/\Omega_{\rm DM}$. Fig.~{\ref{fig7a}} shows such constraints for a null-detection along our fiducial arc when observed with $\calR = 0.8$ mas. The shaded regions mark the 68 and 99~percent confidence regions, and indicate that such a null-detection allows one to rule out $n^*_{\rm BH} \gta 1~\Mpc^{-3}$ for $\alpha \simeq -2$. For a significantly steeper mass function, that is, smaller values of $\alpha$, the constraining power with respect to $n^*_{\rm BH}$ rapidly diminishes. The thick, solid lines correspond to constant values of $\Omega_{\rm BH}/\Omega_{\rm DM}$ of $1$, $0.01$, and $0.0001$, as indicated. 

As is evident, a null-detection along our fiducial curve basically rules out that the majority of dark matter is in the form of black holes characterized by a Schechter-like mass function. The green dot marks the values of $\alpha$ and $n^*_{\rm BH}$ for the mass function of wandering black holes in the {\sc Romulus25} simulation (cf. Fig.~\ref{fig:BHmassfunc}). Note that this estimate is based on the assumption that all SMBHs in {\sc Romulus25} not located at a galactic nucleus are detectable as free-floating black holes. However, roughly 40 percent of them reside within 10 kpc from the central galaxy in the halo in which they orbit. For sufficiently close separations (in projection), it may not be possible, or at least be much harder, to detect the wandering black hole, as the (central) galaxy may dominate the gravitational deflection. We have not attempted to account for this, and therefore caution the reader that the {\sc Romulus25} prediction shown is likely to be somewhat too optimistic. But even in that case, it is clear that being able to rule-out such a prediction requires data that represents an improvement of several orders of magnitude compared to what is achievable with our single, fiducial arc. 

It is useful to translate the above constraints on the number density and power-law slope of the black hole mass function, into constraints on the halo occupation statistics of wandering black holes. Let $\left<N_{\bullet}\right>$ denote the number of wandering black holes with mass $\MBH \geq 10^6\Msun$ in a halo of mass $M_{\rm h}$. Motivated by the occupation statistics of galaxies (see \citet{Wechsler.Tinker.18} for a comprehensive review), we assume that their occupation number scales linearly with halo mass, that is,
\begin{equation}
\langle N_{\bullet} \rangle(M_\rmh) = N_{12} \left(\frac{M_\rmh}{10^{12}\Msun}\right)\,,
\end{equation}
where $N_{12}$ is the {\it average} number of wandering black holes in a halo of mass $M_\rmh = 10^{12} \Msun$.  If we now assume that the dark matter haloes follow the \citet{Sheth.Tormen.02} halo mass function, $n_\rmh(M_\rmh,z) = \rmd n_\rmh / \rmd M_\rmh$, then the number density of wandering black holes at redshift $z$ is given by\\
\begin{equation}\label{nsmbh}
n_{\bullet}(z) = \int_0^\infty \langle N_{\bullet} \rangle(M_\rmh) \, n_\rmh(M_\rmh,z)\, \rmd M_\rmh = \left( \frac{N_{12}}{10^{12}\Msun}\right) \, \bar{\rho}_{\rmm}(z)\, ,
\end{equation}
where we have used that the matter density at redshift $z$ is given by
\begin{equation}
\bar{\rho}_{\rmm}(z) = \int_0^{\infty} M_\rmh \, n_\rmh(M_\rmh,z) \, \rmd M_\rmh\,.
\end{equation}

If we assume that these SMBHs follow the mass function of equation~(\ref{BHmassfunc}), then their number density can also be written as
\begin{equation}\label{nbh}
n_{6+} \equiv \int_{10^6}^{\infty} \frac{\rmd n}{\rmd \MBH} \, \rmd \MBH = n^*_{\rm BH} \, 
\Gamma(\alpha+1,10^{-3})\,.
\end{equation}
Equating $n_{6+}$ with $n_{\bullet}(z=0)$, we obtain that
\begin{equation}
N_{12} = \left( \frac{10^{12} \Msun \, n^*_{\rm BH}}{\rho_{{\rm m},0}}\right) \, \Gamma(\alpha+1,10^{-3})\,.
\end{equation}
We can use this expression to translate our constraints on $n^*_{\rm BH}$ and $\alpha$ into constraints on the halo occupation statistics as characterized by $N_{12}$. The dashed, black contours in Fig.~\ref{fig7a} correspond to contours of fixed $N_{12}$ as labeled. If $\alpha$ is in the range $-2.5 < \alpha < -1$, then a null-detection along our fiducial arc implies that $N_{12} < 10^4$ to $10^5$.  Again, this is far from the occupation numbers predicted by the {\sc Romulus25} simulations, which has $N_{12} = 11.2 \pm 8.4.$\footnote{\citet{Tremmel.etal.18} quote an occupation number of $12.2$, but that includes the SMBH in the centre of the central galaxy.}

\begin{figure*}
\centering
\includegraphics[width=1\textwidth,height=0.5\textwidth]{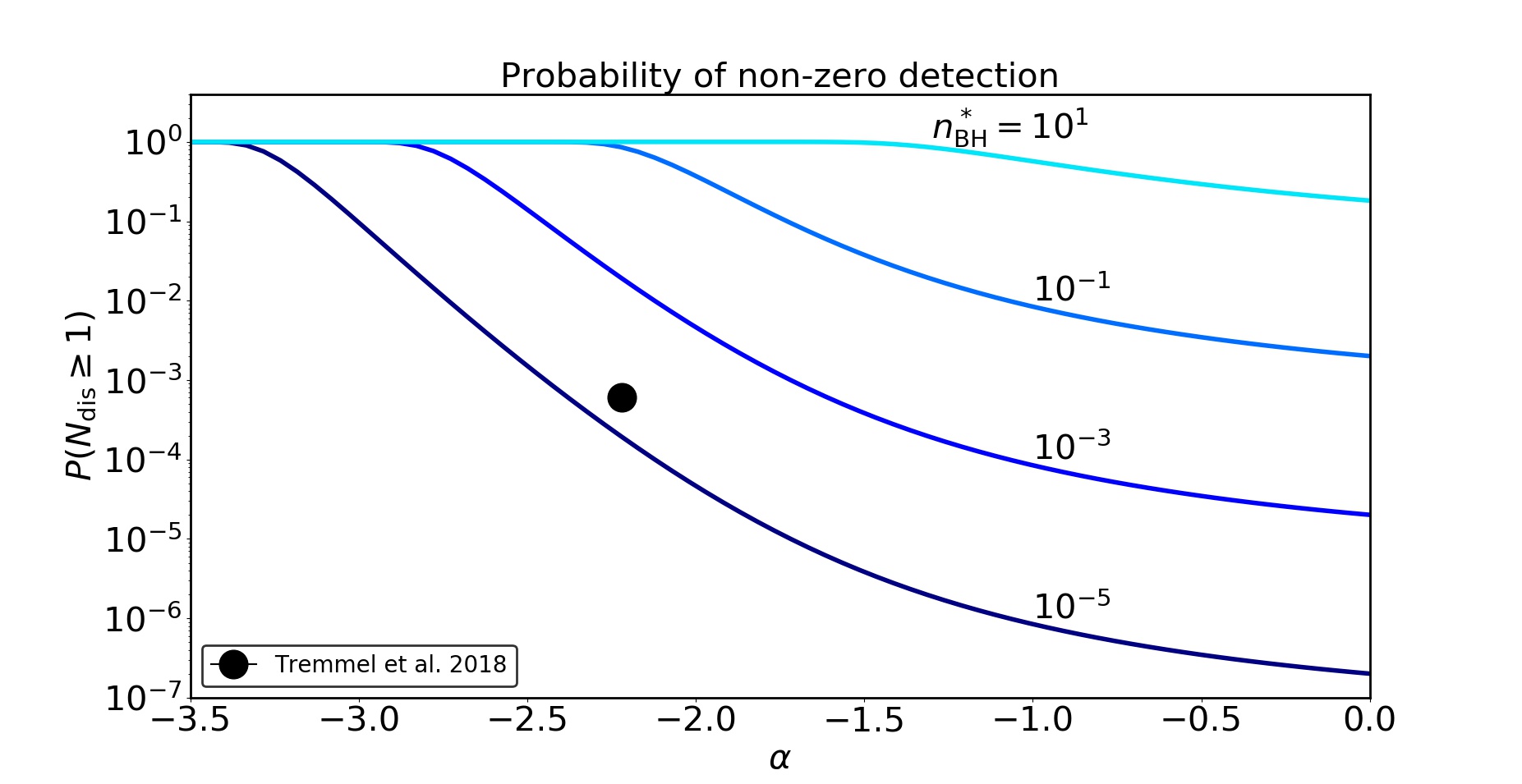}
\caption{Probability of non-zero detection of wandering black holes of mass ranging from $10^{2} \Msun$ to $10^{10} \Msun$ from the distortions of a strong gravitational lensing arc of the fiducial lens system as a function of $\alpha$ for different values of $n^*_{\rm BH}$. The black dot represents the probability of a non-zero detection corresponding to the wandering SMBH occupation number in a Milky Way sized halo and $\alpha$ obtained by \citet{Tremmel.etal.18} in their cosmological simulation {\sc Romulus25}.}
\label{fig7b}
\end{figure*}

Another way to portray these results is to compute, for a given $\alpha$ and $n^*_{\rm BH}$, the probability that one detects at least one distortion along our fiducial arc. Under our assumption of Poisson statistics, this probability is
\begin{equation}
P(N_{\rm dis} \geq 1) = 1-e^{-\lambda_{\rm dis}}\,,
\end{equation}
where $\lambda_{\rm dis}$ is given by equation~(\ref{Nbh}). Results are presented in Fig.~\ref{fig7b}, which shows the probability $P(N_{\rm dis} \geq 1)$ as a function of $\alpha$. Different curves correspond to different values of $n^*_{\rm BH}$, as indicated, and the solid dot indicates the halo occupation prediction of \citet{Tremmel.etal.18}. If their prediction is correct, the probability of detecting at least one distortion along our fiducial arc, with a resolution of $\calR = 0.8$~mas, is only $\sim 10^{-3}$. Put differently, of order $10^3$ such arcs are required before we may reasonably expect to detect a distortion. 

However, as shown in Section~\ref{sec:single} above, being able to observe at a higher spatial resolution, and finding razor-thin arcs at higher redshifts, may bring this number down considerably. As the number of high-resolution, razor-thin lensing arcs continues to increase, so will our ability to improve on these constraints; it remains to be seen whether enough statistics can be accumulated for this test to become meaningful. Future radio surveys with the Square Kilometre Array (SKA) are projected to detect of order $10^5$ gravitational lensed AGN in the radio \citep[][]{McKean.etal.15}. Such sample sizes, coupled with a VLBI capability for the SKA, would turn the method discussed in this paper into an extremely competitive probe of the mass function of free-floating black holes.

\section{Conclusion}
\label{sec:concl}

Razor-thin lensing arcs, with sub-mas width, are ideal probes of the coarseness of the matter distribution along the line of sight towards the source that is being lensed. The thinness of the arcs implies a high sensitivity to detecting lensing distortions due to relatively low-mass objects. In particular, the presence of a free-floating black hole at a sufficiently small impact parameter along the line of sight will cause kink-like distortions in the arc (see Fig.~\ref{illustration}) that will be detectable if the black hole is sufficiently massive. Here free-floating black holes are defined as relatively massive black holes ($M_{\rm BH} \gta 10^2 \Msun$) that are sufficiently separated from any galaxy such that they may be considered as isolated for the purpose of computing the lensing distortions on the razor-thin arcs. They basically fall in two categories: PBHs, that form at early times during the radiation dominated era, and wandering black holes that arise as a consequence of the formation and evolution of SMBHs in galactic nuclei.

In this paper, we have investigated the constraints on the number density of free-floating black holes that one may expect to achieve from observations of kink-like distortions of razor-thin arcs, or a lack thereof. Using the double-lens equation, we compute the extent of these kink-like distortions, which we deem detectable `by eye' (i.e., without any complicated lens modeling of the arc surface brightness distribution) as long as the difference in the perturbations along the radial direction, measured at two different positions along the arc, is larger than half the FWHM of the observation. At a resolution of $\calR \sim 0.8$ mas, which is relatively straightforward to achieve with current VLBI facilities at cm-wavelengths, the minimum mass of a perturbing black hole that is detectable is roughly $10^3 \Msun$. 

We have computed the constraints on the mass and number density of free-floating black holes that are achievable with a fiducial razor-thin arc of length $\arc \sim 270$~mas and an unresolved width equal to the resolution $\calR \sim 0.8$~mas. For our fiducial arc we assume that the source is located at a redshift $z_\rmS = 2.056$, while the main lens is at $z_\rmL = 0.881$. These values are comparable to those of a razor-thin arc observed from the gravitational lens JVAS~B1938+666 with global VLBI at 1.7 GHz (McKean et al. in preparation). If no kink-like distortions are found along such a fiducial arc, one infers that the matter density of free-floating black holes is $\Omega_{\rm BH} < 0.056 \,\Omega_{\rm DM}$ for $M_{\rm BH} \approx 10^6 \Msun$. This is similar to the existing constraint from quasar milli-lensing \citep{Wilkinson.etal.01}. The constraints will improve with increasing total arc length (i.e., more arcs), with increasing source redshift, and, above all, with improved resolution, $\calR$. In fact, the constraint on $\Omega_{\rm BH}/\Omega_{\rm DM}$ scales with $\calR^5$ for black holes with a mass below the characteristic mass $M_0 \simeq 3 \times 10^4 \Msun \, (\calR/{\rm mas})^2 \, (D_{\rmS}/{\rm Gpc})$. For black holes with $M_{\rm BH} \gg M_0$, the constraints have a weak dependence on $\calR$. A null-detection along our fiducial arc, but with a resolution of $\calR = 0.2$~mas, which might be achievable in the near future through observations at a higher frequency, would imply $\Omega_{\rm BH} < 0.013\, \Omega_{\rm DM}$ for $M_{\rm BH} \approx 10^5 \Msun$. 

To put these constraints in perspective, we have examined the demographics of wandering black holes in the state-of-the-art hydrodynamical simulation {\sc Romulus25} \citep[][]{Tremmel.etal.18}, which predicts of order 10 wandering black holes with mass $M_{\rm BH} > 10^6~\Msun$ per Milky-Way like halo. The mass function is well fit by a Schechter function with slope $\alpha \sim -2.2$ and characteristic cut-off mass $M^*_{\rm BH} \sim 10^9~\Msun$. If we assume that the occupation number of wandering black holes scales linearly with halo mass, the implied mass density of wandering black holes with mass $M_{\rm BH} > 10^3 \Msun$ is of order $\Omega_{\rm BH} = 10^{-4}\, \Omega_{\rm DM}$. This is more than three orders of magnitude lower than the constraints achievable with our fiducial arc. Put differently, if the predictions of {\sc Romulus25} are correct, and there is no additional contribution from PBHs, then the probability of detecting a kink-like distortion due to a free-floating black hole along our fiducial arc is only $\sim 10^{-3}$.

Hence, we are left to conclude that razor-thin arcs, observed at sub-mas resolution, can only provide competitive constraints on the mass density of free-floating black holes if the astronomical community is able to amass a large sample of razor-thin arcs, preferentially associated with sources at high redshift, and observed with the highest-possible spatial resolution. On a positive note, it is important to point out that we have only considered kink-like distortions that are detectable `by eye'. When using sophisticated image analysis techniques, similar to what has been used for a number of existing lenses at optical and mm-wavelengths \citep[e.g.][]{Vegetti.etal.10, Vegetti.etal.12, Vegetti.etal.14, Hezaveh.etal.16b}, we expect that one ought to be able to improve sensitivity by at least an order of magnitude, resulting in a similar order of magnitude improvement in the constraints on  $\Omega_{\rm BH}/\Omega_{\rm DM}$. Hence, we remain optimistic that razor-thin arcs will prove to be a powerful probe of the coarseness of the matter distribution on cosmological scales, and of the mass density of free-floating black holes in particular.

\section*{Acknowledgements}

We are grateful to Dhruba Dutta Chowdhury, Johannes Lange, Tim Miller, and Nir Mandelker for help and useful discussions, and to the anonymous referee for valuable comments and suggestions. FvdB is supported by the National Aeronautics and Space Administration under grant no. 17-ATP17-0028 issued through the Astrophysics Theory Program and by the US National Science Foundation through grant AST 1516962 and receives additional support from the Klaus Tschira foundation. FvdB, AM, and SM are grateful to the Kavli Institute for Theoretical Astrophysics at University of California Santa Barbara for support that promoted some of the early discussions regarding this research. SM was supported by Japan Society for the Promotion of Science Kakenhi grant no. 16H01089. This work was supported by World Premier International Research Center Initiative (WPI Initiative), MEXT, Japan.




\bibliographystyle{mnras}
\bibliography{references_vdb}



\appendix

\section{Scaling relations}
\label{sec:scale}

The effective volume specified in equation~(\ref{comovingvol}) has to be computed numerically, and involves numerically solving the perturbative form of the double lens equation to compute $\Delta \beta_{\rm P,max}(z)$. In order to provide some insight as to how $V_{\rm eff}$ scales with the mass of the perturber, $M_\rmP$, and the resolution limit, $\calR$, it is useful to derive some scaling relations by analytically solving the perturbative lens equation under certain extreme conditions.

The perturbative double lens equation given in equation (\ref{matrixnot}) can be analytically solved in certain simplified cases. We shall compute $\delta \vec{\theta}$ in one such simplified foreground case. At the position along the unperturbed arc closest to the perturber (point 1 in Fig.~\ref{angles_illustration}), we have that $\theta_{\rm{0x1}}=0$ and $\theta_{\rm{0y1}} = |\vec{\theta_0}|$. Hence, both the magnification and correction tensor (equations~[\ref{M}] and [\ref{C}]), reduce to diagonal form:
\begin{equation}
\begin{aligned}
\textbf{M} (\vec{\theta_0}) = \begin{bmatrix} 
\left(1 - \theta_\rmE/|\vec{\theta_0}|\right)^{-1} & 0 \\ 
0 & 1
\end{bmatrix}
\end{aligned}
\label{simpleM}
\end{equation}
and
\begin{equation}
\begin{aligned}
\textbf{C} (\vec{\theta_0}) = \begin{bmatrix} 
1-\gamma_\rmf \,\theta_\rmE / |\vec{\theta_0}| & 0 \\ 
0 & 1 
\end{bmatrix}\,.
\end{aligned}
\label{simpleC}
\end{equation}

Solving equation~(\ref{matrixnot}) in the foreground case using the above simplified forms of $\textbf{M}$ and $\textbf{C}$, and the form for $\vec{\alpha_{\rm DP}}$ given in equation~(\ref{perturber}), yields
\begin{equation}
\begin{aligned}
\delta \theta_{\rmx1} &= 0\,,\\
\delta \theta_{\rmy1} &= \frac{{\theta}^2_{\rm EP}}{\delta \theta_{\rmy1} + \Delta \beta_\rmP}\,,
\end{aligned}
\label{delthetar1}
\end{equation}
where $\Delta \beta_\rmP = \theta_{0\rmy1}-\beta_{\rm Py} = |\vec{\theta_0}| - \beta_{\rm Py}$ is the angular impact parameter of the perturber with respect to the unperturbed arc. The solution for $\delta \theta_{\rmy1}$ is similar to the expression for the image angle as a function of the angular impact parameter $\Delta \beta_\rmP$ in absence of the main lens:
\begin{equation}
\delta \theta_{\rmy1} = -\frac{\Delta \beta_\rmP}{2} \pm \sqrt[]{{\left(\frac{\Delta \beta_\rmP}{2}\right)}^2+{\theta}^2_{\rm EP}}\,.
\label{deltathetamax}
\end{equation}
Note that if $\Delta \beta_\rmP=0$, the perturbation of the arc, $\delta \theta_{\rmy1}$ is equal to $\theta_{\rm EP}$, the Einstein angle of the perturber, as expected. Also note that the tangential perturbation, $\delta \theta_{\rmx1}$ is zero and the radial perturbation, $\delta \theta_{\rmy1}$ is maximum at this point. The solution for $\delta \vec{\theta}$ at this point for the background case yields similar results.

The computation of $\delta \vec{\theta}$ at some other point on the arc is more involved because the magnification and correction tensors now have non-zero off-diagonal terms. At point 2, located an angle $f\omega_{\rm arc}/2$ along the unperturbed arc from point 1 (see Fig. \ref{angles_illustration}), we have that $\theta_{\rm{0x2}} = |\vec{\theta_0}| \, {\rm sin}(\phi)$, and $\theta_{\rm{0y2}} = |\vec{\theta_0}| \, {\rm cos}(\phi)$. If $\phi$ is sufficiently small\footnote{This is typically the case if the arc length, $\omega_{\rm arc}$, is only a fraction of the entire length of the Einstein ring, $2 \pi |\vec{\theta_0}|$.}, then, to first order, $\theta_{\rm{0x2}}\approx |\vec{\theta_0}| \, \phi$ and $\theta_{\rm{0y2}} \approx |\vec{\theta_0}|$. Under this approximation, $\textbf{M}$ and $\textbf{C}$ have off-diagonal terms proportional to $\phi$ and diagonal terms independent of $\phi$. Solving for $\delta \vec{\theta}$ using equation~(\ref{matrixnot}) for the foreground case, one can obtain the following form for the radial perturbation $\delta \theta_{\rmr2}$ by applying the rotation matrix as in equation (\ref{radtandeltheta})
\begin{equation}
\begin{aligned}
\delta \theta_{\rmr2} &\approx \phi\,\delta \theta_{\rmx2} + \delta \theta_{\rmy2}\\
&\approx\frac{{\theta}^2_{\rm EP}}{{\left(\delta \theta_{\rmx2} + |\vec{\theta_0}| \, \phi\right)}^2+{\left(\delta \theta_{\rmy2} + \Delta \beta_\rmP \right)}^2}\\
&\times \left[ |\vec{\theta_0}| \, {\phi}^2 + \delta \theta_{\rmr2}+\Delta \beta_\rmP\right],
\end{aligned}
\label{delthetar2}
\end{equation}
where $\Delta \beta_\rmP = \theta_{0\rmy2}-\beta_{\rm Py} \approx |\vec{\theta_0}| - \beta_{\rm Py}$. 

Ignoring the perturbations in the equations (\ref{delthetar1}) and (\ref{delthetar2}) relative to $\Delta \beta_\rmP$ and $|\vec{\theta_0}| \, \phi$, we have the following expressions for $\delta \theta_{\rmr 1}$ and $\delta \theta_{\rmr 2}$
\begin{equation}
\begin{aligned}
\delta \theta_{\rmr 1} &\approx \frac{{\theta}^2_{\rm EP}}{\Delta \beta_\rmP}\,,\\
\delta \theta_{\rmr 2} &\approx \frac{{\theta}^2_{\rm EP}}{{|\vec{\theta_0}|}^2{\phi}^2+{\left(\Delta \beta_\rmP\right)}^2}\left[|\vec{\theta_0}| \, {\phi}^2 + \Delta \beta_\rmP \right]\,.
\end{aligned}
\label{delthetar12}
\end{equation}
Implementing the limiting case of the detectability criterion given in equation (\ref{detectcriteria}), i.e. $\Delta \theta_\rmr = \delta \theta_{\rmr1} - \delta \theta_{\rmr 2} = \calR/2$, we obtain the following equation for $\Delta \beta_{\rm{P,max}}$
\begin{equation}
\begin{aligned}
&{\left(\Delta \beta_{\rm{P,max}}\right)}^3 + \Delta \beta_{\rm{P,max}} \left( |\vec{\theta_0}| + \frac{2\theta^2_{\rm{EP}}}{\calR}\right) \, |\vec{\theta_0}| \, {\phi}^2 \\
&- \frac{2\theta^2_{\rm{EP}}}{\calR} |\vec{\theta_0}|^2 \, \phi^2 = 0\,.
\end{aligned}
\label{deltabetapmaxeqn}
\end{equation}
In the background case, $\Delta \beta_{\rm P,max}$ follows a similar equation, but with $|\vec{\theta'_0}| = |\vec{\theta_0}| - \gamma_\rmb \, \theta_\rmE$ replacing $|\vec{\theta_0}|$.

It can be seen from equation~(\ref{deltabetapmaxeqn}) that the product of the roots, $(2 \theta^2_{\rm{EP}}/\calR) \, |\vec{\theta_0}|^2 \, \phi^2$, is positive definite, whereas the sum of the roots, i.e., the coefficient of the quadratic term, is equal to zero. This implies that the equation has only one positive real root. As we demonstrate below, this root behaves differently for small and
large $z_\rmP$.

We can also derive an approximate expression for $M_{\rm P,min}$, the minimum BH mass that can be detected at a given redshift. The maximum distortion is caused when the impact parameter is equal to zero, in which case the radius of the kink-like distortion, $\delta \theta_{\rmy1}$, at point 1 (see Fig. \ref{angles_illustration}), is equal to the Einstein radius of the perturbing black hole. Hence, based on our detectability criterion (equation~[\ref{detectcriteria}]), $M_{\rm P,min}$ corresponds to the mass for which $\theta_{\rm EP} = \calR/2$. This translates to 
\begin{equation}
M_{\rm P,min}(z_\rmP) = \frac{\calR^2c^2}{16G} \frac{D_{\rmP} D_{\rmS}}{D_{\rm{PS}}} \equiv M_0 \, \frac{D_{\rmP}}{D_{\rm PS}}\,,
\label {minmass}
\end{equation}
where we have defined a the characteristic mass-scale
\begin{equation}\label{M0char}
M_0 \equiv \frac{\calR^2 \, c^2}{16\, G} \, D_{\rmS}\,.
\end{equation}
For our fiducial case, which has $z_\rmS=2.056$ and $\calR=0.8$ mas, we have that $M_0 \approx 3.5\times10^4 \Msun$. 

In what follows we use equations (\ref{deltabetapmaxeqn}) and (\ref{minmass}) to derive approximate expressions for the effective volume, $V_{\rm eff}$, given by equation~(\ref{comovingvol}) in the limits of low mass ($M_\rmP \ll M_0$) and high mass ($M_\rmP \gg M_0$). 

\subsection{Low mass limit}

In the low mass limit, i.e. $M_\rmP \ll M_{0}$, the maximum redshift out to which the perturber can be detected, $z_{\rm P,max}$ is small, and the effective comoving volume, $V_{\rm eff}$, comprises only of the foreground volume. In addition, we have that $D_\rma(z_{\rm P,max}) \approx r(z_{\rm P,max}) \approx (c/H_0) z_{\rm P,max}$. The comoving volume element is given by
\begin{equation}
\frac{\rmd^2 V}{\rmd\Omega \, \rmd z} = D_\rmH \, \frac{r^2(z)}{E(z)}
\label{dcomovingvoldOmegadz}
\end{equation}
with $D_\rmH = c/H_0$,  $r(z)$ the comoving distance out to redshift $z$, and $E(z) = H(z)/H_0$. At small $z$, we have that $E(z) \sim 1$, and the expression for the effective comoving volume (eq.~[\ref{comovingvol}]) reduces to
\begin{equation}
V_{\rm eff} \approx 2 \, \arc \, D^3_\rmH \, \int\limits_0^{z_{\rm P,max}}
\Delta\beta_{\rm P,max}(z) \, z^2 \, \rmd z\,.
\label{Veffsmall}
\end{equation}

The maximum redshift, $z_{\rm P,max}$, is the redshift at which the perturber has an Einstein radius $\theta_{\rm EP} = \calR/2$, and thus where $M_\rmP = M_{\rm P,min}$.
For sufficiently small $M_\rmP$, this allows us to write
\begin{equation}\label{zPmax}
z_{\rm P,max} \approx \frac{M_\rmP}{M_0} \, \frac{D_{\rmS}}{D_\rmH}\,.
\end{equation}

For $z_\rmP \ll z_{\rm P,max}$, ${\left(\Delta \beta_{\rm P,max}\right)}^3$ is the dominant term in equation (\ref{deltabetapmaxeqn}) and therefore in this limit, $\Delta \beta_{\rm P,max} \approx {\left[\left(2{\theta}^2_{\rm EP}{|\vec{\theta_0}|}^2{\phi}^2\right)/\calR\right]}^{\frac{1}{3}}$. But for most of the redshift range the second term dominates the first in equation~(\ref{deltabetapmaxeqn}), and
\begin{equation}
\Delta \beta_{\rm P,max} \approx \left(\frac{2\theta^2_{\rm EP}}{\calR}|\vec{\theta_0}|\right)\bigg/ \left(|\vec{\theta_0}| + \frac{2\theta^2_{\rm EP}}{\calR}\right) \approx \frac{2\theta^2_{\rm EP}}{\calR}\,,
\end{equation}
where the second approximation follows from the fact that in the small mass limit $|\vec{\theta_0}| \gg \left(2\theta^2_{\rm EP}\right)/\calR$. Substituting this value for $\Delta \beta_{\rm P,max}$ in equation~(\ref{Veffsmall}), we finally obtain that
\begin{equation}
\begin{aligned}
V_{\mathrm{eff}}(M_\rmP \ll M_{0}) \approx 256 \, \arc \, \frac{R_\rms^3}{\calR^5}\,.
\label{Veffscalesmall}
\end{aligned}
\end{equation}

Therefore in the low mass limit the effective volume scales as 
\begin{equation}
V_{\mathrm{eff}}(M_\rmP \ll M_{0}) \sim \arc \, \frac{M^3_\rmP}{\calR^5}\,.
\end{equation}

\subsection{Large mass limit}

Next we consider the large mass limit, where $M_\rmP \gg M_0$. Note, though, that we demand that $M_\rmP$ remains small compared to the mass of the main lens, $M_\rmL$. When $M_\rmP$ is large, we have that $z_{\rm P,max} = z_\rmS$ and the effective comoving volume therefore comprises of both the foreground and background volumes:
\begin{equation}
V_{\rm eff} = 2 \, \arc \int\limits_0^{z_\rmS} \Delta\beta_{\rm P,max}(z) \,
\frac{r^2(z)}{E(z)} \rmd z\,.
\label{Vefflarge}
\end{equation}

In this limit, for most of the redshift range, ${\left(\Delta \beta_{\rm P,max}\right)}^3$ is the dominant term in equation (\ref{deltabetapmaxeqn}), implying that 
\begin{align}
\Delta \beta_{\rm P,max} \approx \begin{cases}
\left[\frac{2{\theta}^2_{\rm EP}}{\calR} \, |\vec{\theta_0}|^2 \, \phi^2 \right]^{1/3}\,, & \text{foreground,} \\
\left[\frac{2{\theta}^2_{\rm EP}}{\calR} \, |\vec{\theta'_0}|^2 \, \phi^2 \right]^{1/3}\,, & \text{background,} \\
\end{cases}
\end{align}
where $|\vec{\theta'_0}| = |\vec{\theta_0}| - \gamma_\rmb \theta_\rmE$. Substituting in equation (\ref{Vefflarge}), and using that $\phi = (f\arc)/(2|\vec{\theta_0}|)$, we finally obtain that in the large mass limit the effective volume scales as
\begin{equation}
V_{\rm eff}(M_\rmP \gg M_{0}) \sim {\left(\arc\right)}^{\frac{5}{3}} f^{\frac{2}{3}} {\left(\frac{M_\rmP}{\calR}\right)}^{\frac{1}{3}}\,.
\label{Veffscalelarge}
\end{equation}

From the two different scaling dependences of $V_{\rm eff}$ in the two limits, one can define a turnover mass $M_{\rm TO}$ by equating the expressions for $V_{\rm eff}$ in the low and high mass limits. This yields
\begin{equation}
\begin{aligned}
M_{\rm TO} &\approx 4\times 10^5 {\left(\frac{f}{0.5}\right)}^{\frac{1}{4}} {\left(\frac{\arc}{270 \,\text{mas}}\right)}^{\frac{1}{4}} {\left(\frac{\calR}{0.8 \,\text{mas}}\right)}^{\frac{7}{4}}\\
&\times {\left(\frac{r(z_\rmS)}{1 \,\text{Gpc}}\right)}^{\frac{5}{8}} \Msun\,,
\end{aligned}
\label{turnovermass}
\end{equation}
which is about an order of magnitude larger than the characteristic mass $M_0$. Note that $M_{\rm TO}$ 
characterizes the mass at which the mass-scaling of the effective volume changes its behavior, and roughly corresponds to the mass at which the constraints on $\Omega_{\rm {BH}}/\Omega_{\rm {DM}}$ are most strict
(cf. Fig.~\ref{fig6}).

\section{Distortions due to NFW haloes}
\label{sec:subhalo}

In this paper, we examine how {\it localized} kink-like distortions of gravitational arcs can be used to constrain the number density of free-floating black holes. However, one might envision other objects to cause similar distortions. In particular, in the $\Lambda$CDM paradigm one expects a large abundance of dark matter haloes spanning many orders of magnitude in halo mass. These haloes are predicted to have a universal Navarro-Frenk-White (NFW) density profile \citep{Navarro.etal.97}, and each halo in turn is predicted to have a hierarchy of substructures \citep[e.g.][]{Jiang.vdBosch.16}. One might worry that such dark matter (sub)haloes along the line of sight cause similar distortions of razor-thin lensing arcs, and thereby `confuse' our interpretation in terms of free-floating black holes. After all, several groups are using distortions of gravitational lenses to probe dark matter (sub)-structure \citep[e.g.,][]{Koopmans.05, More.etal.09, Vegetti.Koopmans.09, Vegetti.etal.10, Vegetti.etal.12, Hezaveh.etal.16b, Birrer.etal.17}. However, because of their wildly different density profiles, the lensing distortions caused by black holes cannot be confused for distortions caused by NFW haloes.

\begin{figure}
\hspace{-0.25cm}
\includegraphics[width=0.5\textwidth,height=0.55\textwidth]{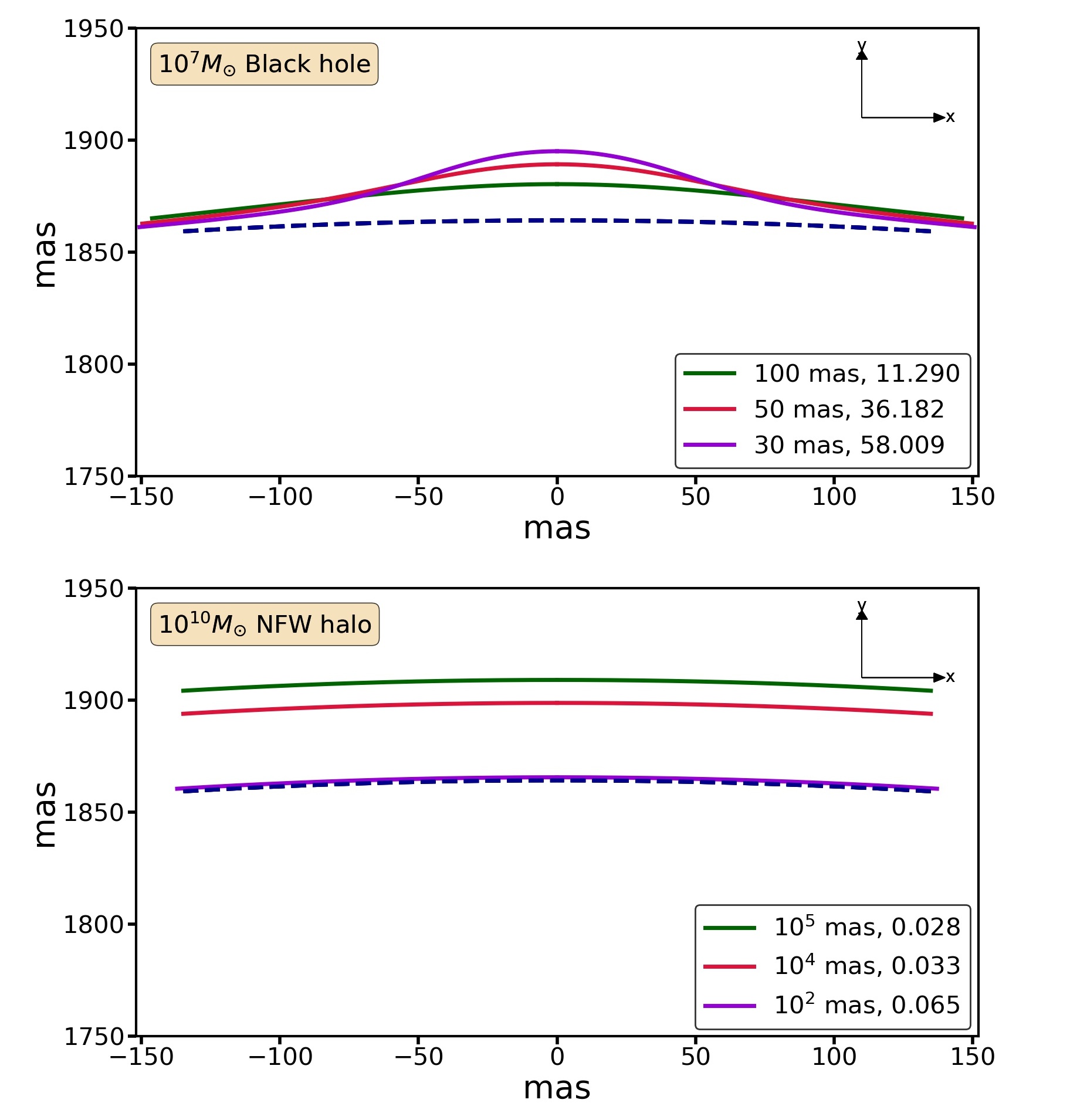}
\caption{The upper panel shows the perturbed arcs due to a $10^7 \Msun$ black hole at $z_\rmP=0.01$ for three different values of the angular impact parameter $\Delta \beta_\rmP$. Here we consider our fiducial lensing configuration with $z_\rmL=0.881$, $z_\rmS=2.056$, $M_\rmL(\theta_\rmE) = 10^{12} \Msun$ and $|\vec{\beta_\rmS}|=36$ mas. The lower panel shows the perturbed arcs in the presence of an NFW halo perturber with $M_{\rm vir}=10^{10} \Msun$ and $c_{\rm vir}=10$. The corresponding values of $2 \Delta \theta_\rmr/\calR$ (assuming the fiducial value of $\calR=0.8$ mas and $f=0.5$) are also indicated. If this ratio is larger than unity the perturbation is deemed detectable according to our criterion (equation [\ref{detectcriteria}]). As is evident, whereas the black hole perturber is detectable for all three impact parameters shown, the distortions due to the NFW halo are not sufficiently localized to pass our detection criterion.}
\label{subhalo_distortion}
\end{figure}

To illustrate this, Fig.~\ref{subhalo_distortion} compares the distortions due to a black hole of mass $M_\rmP=10^7 \Msun$ (upper panel) to those due to an NFW halo of mass $M_{\rm vir}=10^{10} \Msun$ and with a concentration parameter $c_{\rm vir}=10$ (lower panel). These have been computed by numerically solving the perturbative form of the double lens equation (\ref{matrixnot}). In each case we show the results for three different angular impact parameters, and we also indicate the value of $2 \Delta \theta_\rmr/\calR$; this ratio needs to be larger than unity for the perturbation to be considered `detectable' according to our criterion (equation~[\ref{detectcriteria}]). In the case of the black hole, the perturbations easily satisfy our detection criterion, for all three values of the angular impact parameter. In the case of the NFW halo, though, $2 \Delta \theta_\rmr/\calR < 0.1$ for all cases shown, and these distortions are therefore not `detectable' based on our criterion. The reason is that the distortions are not sufficiently localized. Rather, the arc is merely displaced from its unperturbed location. An NFW perturber causes a maximal displacement of the arc when the impact parameter with respect to the unperturbed arc is approximately equal to the halo's scale radius. For the halo considered here, the scale radius is $r_\rms \approx 7.5$ kpc, which corresponds to an angular impact parameter of $\approx 3.5 \times 10^4$ mas at the redshift of the perturber (here taken to be $z_\rmP = 0.01$). Note that the NFW perturber has a mass that is 1000 times larger that that of the black hole; if we were to consider an NFW perturber with a mass of $10^7 \Msun$, the `displacement' of the arc would be a factor of 100 smaller, and the maximum value of $2 \Delta \theta_\rmr/\calR$ would be only $\sim 0.008$. Hence, under no circumstance can NFW haloes along the line of sight cause localized, kink-like distortions such as those due to black hole perturbers.

\bsp	
\label{lastpage}
\end{document}